\documentclass{emulateapj}
\usepackage{epsfig}
\usepackage{lscape}
\usepackage{amssymb,amsmath}

\def\lx{L$_x$}
\def\rc{r$_c$}
\def\rh{r$_h$}

\def\x{$\times$}
\def\about{$\sim$}
\def\simlt{$\la$}
\def\simgt{$\ga$}

\def\asec{$''$}
\def\amin{$'$}
\def\spt{$\buildrel{\mathrm{s}}\over .$} 
\def\secspt{$\buildrel{\prime\prime}\over .$}
\def\minspt{$\buildrel{\prime}\over .$}

\def\mass{${\cal M}$}
\def\msun{${\cal M}_{\odot}$}

\def\yr-1{yr$^{-1}$}
\def\b{$B_{435}$}
\def\R625{$R_{625}$}
\def\br{$B_{435}-R_{625}$}
\def\hr{$H\alpha-R_{625}$}
\def\ha{H$\alpha$}
\def\hb{H$\beta$}
\def\vi{$V_{555}-I_{814}$}
\def\sig{$\sigma$}

\def\Teff{T$_{eff}$}

\begin{document}
\title{Helium-Core White Dwarfs in the Globular Cluster NGC 6397}
\author{Rachel R. Strickler\altaffilmark{1,2}, 
Adrienne M. Cool\altaffilmark{2},
Jay Anderson\altaffilmark{3},  
Haldan N. Cohn\altaffilmark{4},
Phyllis M. Lugger\altaffilmark{4},
Aldo M. Serenelli\altaffilmark{5} 
\\ \today
}
\altaffiltext{1}{Department of Astronomy and Astrophysics, University of California, Santa Cruz,
287 Interdisciplinary Sciences Building (ISB), Santa Cruz, CA 95064; rstrickl@ucsc.edu}
\altaffiltext{2}{Department of Physics and Astronomy, San Francisco State University,
1600 Holloway Avenue, San Francisco, CA 94132; cool@sfsu.edu}
\altaffiltext{3}{Space Telescope Science Institute, Baltimore, MD 21218; jayander@stsci.edu} 
\altaffiltext{4}{Department of Astronomy, Indiana University, Swain West, Bloomington, IN 47405; 
cohn@astro.indiana.edu, lugger@astro.indiana.edu} 
\altaffiltext{5}{Max-Planck-Institut f\"ur Astrophysik,
  Karl-Schwarzschild-Str. 1, Garching, 85471,
  Germany; aldos@mpa-garching.mpg.de}

\begin{abstract} 

We present results of a study of the central regions of NGC~6397 using
Hubble Space Telescope's Advanced Camera for Surveys (HST ACS),
focusing on a group of 24 faint blue stars that form a sequence
parallel to, but brighter than, the more populated sequence of
carbon-oxygen white dwarfs (CO~WDs).  Using F625W, F435W, and F658N
filters with the Wide Field Channel (WFC) we show that these stars, 18
of which are newly discovered, have magnitudes and colors consistent
with those of helium-core white dwarfs (He~WDs) with masses \about
$0.2-0.3$\msun.  Their \hr\ colors indicate that they have strong \ha\
absorption lines, which distinguishes them from cataclysmic variables
in the cluster.  The radial distribution of the He~WDs is
significantly more concentrated to the cluster center than that of
either the CO~WDs or the turnoff stars and most closely resembles that
of the cluster's blue stragglers.  Binary companions are required to
explain the implied dynamical masses.  We show that the companions
cannot be main-sequence stars and are most likely heavy CO~WDs.  The
number and photometric masses of the observed He~WDs can be understood
if \about $1-5$\% of the main-sequence stars within the half-mass
radius of the cluster have white dwarf companions with orbital periods
in the range \about $1-20$ days at the time they reach the turnoff.
In contrast to the CO~WD sequence, the He~WD sequence comes to an end
at \R625\ $\simeq$ 24.5, well above the magnitude limit of the
observations.  We explore the significance of this finding in the
context of thick vs.\ thin hydrogen envelope models and compare our
results to existing theoretical predictions.  In addition, we find
strong evidence that the vast majority of the CO~WDs in NGC~6397 down
to \Teff\ $\simeq$ 10,000~K are of the DA class.  Finally, we use the
CO~WD sequence together with theoretical cooling models to measure a
distance to the cluster of 2.34~$\pm$~0.13~kpc, or $(m-M)_{625} =
12.33$.
\end{abstract}


\section{Introduction} \label{section:intro}


Hubble Space Telescope (HST) has opened a new era in the study of
white dwarfs by making it possible to study these stars within
globular clusters for the first time.  Early HST observations revealed
handfuls of white dwarfs in several clusters; since then, extensive
cooling sequences comprising hundreds of white dwarfs have been
observed in nearby clusters---e.g., \citet{renzini96, cpk96, richer97,
zoccali01, hansen04, hansen07, calamida08}.  The advantages of
studying a population of stars all at the same, well-known distance
can thus be brought to bear on the study of the final stage of
evolution of low-mass stars.

As the second-nearest globular cluster to the Sun, and one with a
well-constrained reddening \citep{gratton03}, NGC~6397 is particularly
favorable for the study of white dwarfs (WDs).  Early HST Wide Field
Planetary Camera 2 (WFPC2) observations revealed a narrow sequence of
stars whose position was consistent with carbon-oxygen WDs (CO~WDs) of
a single mass in the range \about 0.5-0.6\msun\ \citep{cpk96}.  Most
recently, \citet{hansen07} have measured the distance and age of the
cluster by detecting the bottom of the WD sequence, and
\citet{davis08} have found evidence for natal kicks among CO~WDs in
the cluster.  Both studies make use of deep HST/ACS imaging of a field
outside the half-mass radius \citep{richer08}.

NGC~6397 is also the nearest apparently core-collapsed globular
cluster \citep{tkd95}.  It has a small, resolved core of uncertain
radius ($1.5-8$\asec\ at 95\% confidence---\citet{sosin97}) surrounded
by a power-law ``cusp'' region extending out to \about 100\asec\
\citep{loam92, lcg95}.  A wide variety of objects whose origins are
thought to be linked to its high central density and resulting high
rate of stellar interactions have been discovered in this central cusp
region.  These include numerous blue stragglers \citep{alo90},
cataclysmic variables (CVs) \citep{cgc95, ghe01, cohn09}, a quiescent
low-mass X-ray binary \citep{ghe01}, and a millisecond pulsar (MSP)
\citep{dpm01}.

A recent addition to the growing list of stellar exotica in NGC~6397
is a set of faint blue stars thought to be low-mass white dwarfs with
helium cores (He~WDs).  The first three of these were found
serendipitously in a search for CVs using HST/WFPC2 \citep{cgc98}.
Dubbed ``non-flickerers,'' they showed neither the short-term
variability nor the UV excess typical of CVs.  Three more candidates
were found in a subsequent WFPC2 study that included \ha\ imaging; all
six candidates were found to be \ha-faint, consistent with their
proposed classification as He~WDs \citep{tge01}.  The one candidate
for which a spectrum has been obtained (NF2) revealed a broad \hb\
line that confirmed its identity as a low-mass white dwarf
\citep{egc99}.

Helium-core WDs have the potential to provide new insight into a
cluster's population of binary stars---stars that play a critical role
in globular cluster dynamics \citep{hut92}.  
The six He~WD candidates
identified in NGC~6397 as of 2001 are confined to the central \about
30\asec\ of the cluster, a distribution that strongly suggests the
presence of unseen binary companions \citep{cgc98, tge01}.  The
spectrum of NF2, with its apparently Doppler-shifted \hb\ line, also
hints at the influence of a dark, compact companion \citep{egc99}.  
Double-degenerate He WD binaries are of additional interest as 
potential progenitors of AM~CVn stars \citep{nelemans01b} and sources
of gravitational radiation \citep{benacquista06}.

In the field, He~WDs are commonly found in binary systems, typically
in ultra-compact X-ray binaries (ULXBs; \citet{njs06} and references therein), 
as companions to MSPs \citep{vk05} or in double-WD systems \citep{mdd95,
mmm02}.  In globular clusters, eight He~WDs have been identified other than
those in NGC~6397.  Three are in ultra-compact X-ray binaries---in
NGC~6712 \citep{and93}, NGC~6624 \citep{and97} and M15
\citep{dieball05}.  Three are companions to MSPs---in 47~Tuc
\citep{egh01}, NGC~6752 \citep{fps03} and M4 \citep{srh03}.  In M4 a
second He~WD has been found as the companion to a subdwarf B star
\citep{otoole06}, while in 47~Tuc a second was identified recently
in a spectroscopic study of blue stars \citep{knigge08}.

From an evolutionary point of view, the connection to binarity follows
from the need to prevent helium ignition at the tip of the red giant
branch (RGB)---otherwise a CO~WD will result, as for single stars.  If
a giant's envelope is removed before helium ignition, then the
degenerate core will be exposed and produce a low-mass
white dwarf with a helium core.  Removal of the envelope may be
accomplished either via Roche-lobe overflow onto a pre-existing close
binary companion \citep{webbink75} or a collision between an RGB star
and a compact object, leading to a common envelope phase followed by
envelope ejection \citep{dbh91}.  Either way, the He~WD is left in a
binary.  He~WDs can also form from single stars if they can somehow
lose an unusually large amount of mass while on the RGB \citep{cc93}.
However, as we will show, the radial distribution of the He~WDs in NGC
6397 strongly suggests that they have binary companions, making such
scenarios unlikely to apply in this case.

Here we use the Advanced Camera for Surveys' Wide Field Channel
(ACS/WFC) to study the central regions of NGC~6397 in more detail.
Our primary focus in the present paper is on the He~WDs.  With the
wider field of view, improved spatial resolution, and greater
sensitivity of ACS/WFC relative to WFPC2, we probe a larger region of
the cluster to fainter magnitudes, and uncover a much larger
population of He~WDs than was previously known.  Future papers will
provide a complete analysis of the full data set and report on our
search for optical counterparts of Chandra X-ray sources in these data
\citep[in preparation]{cohn09}.

Our use of the \ha\ and \R625\ filters also enables us to test for the
presence or absence of an \ha\ absorption line in the atmospheres of
the CO~WDs in the WFC field and thus to classify them as either of the
DA (hydrogen atmosphere) or non-DA type.  The relative numbers of DA
vs.\ non-DA WDs in globular clusters is of significant interest in
view of recent findings that the incidence of non-DA stars appears to
be much lower in open clusters than in the field \citep{krh05}, and
outstanding questions about the mechanisms that give rise to these
differences (\citet{mb08} and references therein).  The possibility
that the deficit of non-DA stars in open clusters may be explained as
a result of the larger masses of open-cluster WDs \citep{krh05} can be
tested by observing globular-cluster WDs, whose masses are similar to
those of field WDs.  All WDs for which classifications have been made
in globular clusters have been found to be DAs \citep{moehler00,
moehler04}.  However the number classified to date is small (4-5 in
each of two clusters) owing to the difficulty of spectroscopic
observations of such faint stars in crowded fields.

Finally, we determine a distance to NGC~6397 using \about 300 of the
best-measured CO~WDs, together with a CO~WD model cooling track kindly
provided by and transformed to the HST/ACS filter system by
P.~Bergeron \citep*{bwb95}.

We describe the data set, together with the photometric analysis and
calibration method, in \S 2.  In \S 3 we identify He~WD candidates
using color-magnitude diagrams (CMDs).  We then consider other
possible explanations for stars appearing in the parts of the CMDs
occupied by the He~WD candidates.  In \S 4 we compare the CO~WDs to
theoretical cooling tracks in a \hr\ vs.~\R625\ CMD to distinguish DA
from non-DA WDs, and discuss the significance of our finding that
most, and perhaps all, of the CO~WDs are of the DA class.  In \S 5 we
present our distance determination.  In \S 6, we analyze the spatial,
mass, and age distributions of the He~WDs, and use the results to
glean information about the nature of their binary companions, the
orbital periods of their progenitor binaries, and the binary fraction
in the cluster.  We also compare our results to predictions made by
\citet*{hkr03} regarding the He~WDs in NGC~6397.  We summarize our
results in \S 7.


\section{Observations and Photometric Analysis} \label{section:obs}


We obtained images using the ACS/WFC comprising a total of five short
and five long exposures each in the F435W and F625W filters (hereafter
\b\ and \R625) and 40 long exposures in the F658N filter (hereafter
\ha). Ten single-orbit visits were made over the course of a year
beginning 2004 July 16.  During each visit, one short and one long
exposure were taken in either \b\ (13~s and 340~s) or \R625\ (10~s and
340~s), followed by four 390~s or 395~s exposures in \ha.  Due to the
changing roll angle of the telescope over the course of the year, a
mosaic constructed from all of the images in a given filter results in
the pinwheel shape seen in Figure 1.  The mosaic provides complete
coverage out to a radius of \about 1\minspt 5 from the cluster center,
partial coverage in two or more exposures out to \about 2\minspt 5, and
partial coverage in at least one exposure out to \about 2\minspt 9.
For comparison, the half-mass radius of NGC~6397 is 2\minspt 33
\citep{harris96}.

Photometry was carried out using the techniques developed by
J.~Anderson that use an ``effective'' point-spread function (PSF)
approach to model how the PSF varies with position on the ACS/WFC
detector.  These methods are described by \citet{ak00} and applied to
the WFC by \citet{ak06}.  We constructed effective PSFs for each of
the three filters from the present data set using a subset of bright,
well-measured stars in each filter.

To generate a star list that would be close to complete while also
excluding most artifacts, we began by finding peaks in each individual
FLT-format image.  We then transformed all coordinates to a common
reference frame and looked for matches within 0.5 pixels.  In \b\ and
\R625, objects that created significant peaks in at least three of the
five images were considered potential stars.  In \ha\ a larger number
of matches was required, depending on the number of images in which
the star could potentially have appeared (up to 40).  Multiple passes
of star-finding were made; during each pass, the stars found in the
previous pass were removed, and additional stars were searched for in
the subtracted images.  When finding fainter stars, we used the method
described by \citet{a08} to avoid false detections due to the `knotty'
PSFs and diffraction spikes associated with bright stars.  The final
collated list contained 25,004 stars.

For stars with \R625\ $<$ 18.5, we measured magnitudes in each
individual frame and then averaged together the results.  This worked
well for bright stars which make strong peaks in all images in which
they appear.  For stars bright enough to saturate in all the long
exposures we adopted magnitudes derived from averages of the short
exposures instead.  So as to obtain a complete census of bright stars,
magnitudes were measured even for stars that are saturated in the
short exposures.  Saturation in the short exposures reduces the
accuracy of the magnitudes measured for stars with \R625\ \simlt\ 14
(see Fig.~2).

For stars with \R625\ $>$ 18.5, we combined the information from all
available frames to measure a single magnitude for a given star.  This
produced noticeably better results for faint stars that do not
necessarily produce strong peaks in every frame.  By incorporating
information from frames in which a given star would otherwise not have
been measured, this method eliminates the bias toward brighter
magnitudes that would otherwise occur for faint stars.

To calibrate the photometry we first selected one long-exposure FLT
image in each filter and performed pixel area map corrections.  Next
we used DAOPHOT \citep{stetson87,stetson94} to measure magnitudes in
0$\farcs$5-radius apertures for 15-20 bright, unsaturated, relatively
isolated stars in each filter.  We converted the resulting aperture
magnitudes to the VEGAMAG system using the prescriptions provided by
\citet{sirianni05}.  We then found the offsets between these
calibrated magnitudes and our instrumental magnitudes for the same
stars.  Applying these offsets to all the instrumental magnitudes then
yielded calibrated magnitudes for the entire data set on the VEGAMAG
system.


\section{An Extended Sequence of Helium-Core White Dwarfs} \label{section:hewdseq}


In Figure 2, we show \R625\ vs.\ \br\ and \R625\ vs.\ \hr\
color-magnitude diagrams (CMDs) for the central regions of NGC~6397.
Here we plot only stars that were detected in at least two frames in
each filter; 15,842 stars made this cut.  We further required that, in
each filter, a star's profile be reasonably well matched to the PSF at
its location using the ``$q$'' quality-of-fit parameter described by
\citet{a08}---see \citet{stricklerMS} for details.  This eliminated
another 866 objects, leaving a final list of 14,976 stars that appear
in these CMDs.  These are the stars that we analyze in the remainder
of this paper.  For stars brighter than \R625\ = 18.5 we plot averages
of magnitudes measured in individual frames.  For all fainter stars we
plot magnitudes derived from simultaneous fitting. 

The most prominent features in these CMDs are the main sequence (MS)
and the RGB, both of which appear nearly vertical in \hr.  NGC 6397's
short blue horizontal branch, blue stragglers, and WD sequence can
also be seen.  In the right panel, these stars all appear to the right
of the RGB or MS, consistent with the stronger \ha\ absorption lines
associated with hotter stars.  In \br\ vs.\ \R625, the majority of the
stars immediately to the left of the MS are likely to be from the
halo/bulge background, while stars to the right of the MS include both
foreground stars and MS-MS binaries (cf. \citet{cb02}).

Taking a closer look at the WDs in the left panel, we see that they
form a densely populated sequence running from \R625\ \about\ 22, \br\
\about\ 0 down to \R625\ \simgt\ 26, \br\ \simgt\ 1.  These are the
carbon-oxygen (CO) WDs associated with single-star evolution in the
cluster---stars that have turned off the main sequence within the past
\about 2 Gyr (cf. \citet{{cpk96}}, \citet{hansen07}).  In addition,
one can see a group of objects that, for a given \br\ color, lie
\about 1-2 magnitudes above the CO~WD sequence.  These stars are well
separated in color from the halo stars, most of which have \br\ $>$ 1,
and include the 6 stars previously identified as probable He~WDs (see
Fig.~3; \citet{cgc98}, \citet{egc99}, \citet{tge01}).  Many more stars
can now be seen in this region of the diagram.  These stars are the
primary focus of the remainder of this paper.

To distinguish potential He~WDs from probable CO~WDs we used the
following procedure.  We note at the outset that we do not expect to
achieve a perfect separation between the two populations.  Our goal is
to select a set of the best He~WD candidates to study in detail while
also retaining for consideration marginal candidates that are less
readily distinguished from the CO~WDs.  To this end we began by
interpolating a 0.53\msun\ CO~WD cooling track from the \citet{bwb95}
models and fitting it to a subset of the WDs chosen
to be very likely CO~WDs.  Details of the fitting procedure are
presented in \S 5.  A best fit was obtained for a distance of
2.34~kpc; the resulting placement of the theoretical cooling curve is
shown in the close-up of the WD region (Fig.~3).

Next we selected all stars within $\pm$ 0.8 magnitudes in \br\ color
of the theoretical CO~WD cooling track, a region which should enclose
all possible WDs while excluding halo stars.  For each star in this
region we determined the offset between its \br\ color and the \br\
color of the 0.53\msun\ cooling track at the same \R625\ magnitude.
We then divided the stars into six bins: one bin for stars from \R625\
= $18.9-23.5$ and five bins 0.5 magnitudes wide from \R625\ = 23.5 to
26.0.  This division produced bins with \about 70-100 stars each.
Next we computed the standard deviation of the color offsets within
each bin.  To minimize the impact of outliers (e.g., He~WDs) on our
determination of \sig\ (which we wished to represent \sig\ for the
CO~WDs exclusively), we iterated 2-5 times (until convergence),
eliminating stars with color offsets greater than 2.5\sig\ at the end
of each iteration.  For a Gaussian distribution of CO~WD colors, this
choice would eliminate \about 1 CO~WD in each bin.  Finally, we
defined a preliminary list of the best He~WD candidates to be stars
that landed more than 3\sig\ away from the cooling track on the red
side.

Two additional criteria were imposed before a star was put on the list
of best He~WD candidates.  First, we required that it be brighter than
\R625\ = 24.5, since measurement uncertainties increase rapidly for
magnitudes fainter than this; the possibility that additional He~WDs
may lie at fainter magnitudes is considered in detail below.  Second,
we required that it be at least 0.75 mag brighter than the 0.53\msun\
cooling sequence to eliminate possible double CO-WDs; this last
requirement eliminated two additional stars.  The 24 stars that met
all these criteria are marked with large open circles in Fig.~3.  The
six previously identified candidates \citep{cgc98, tge01} are among
these and are shaded grey.  These 24 He~WD candidates are listed in
order of increasing \R625\ magnitude in Table~1.  For each star we
provide, in columns 1-8, an ID number, the x and y coordinate in our
\R625\ mosaic, the R.A. and Dec. in J2000 coordinates, the radial
offset from the cluster center in arcseconds, the \R625\ magnitude,
the \br\ color, and the \hr\ color.

To identify probable CO~WDs, we followed the same procedure but used a
more stringent requirement in doing the sigma clipping (2.25\sig\
instead of 2.5\sig), which yielded slightly smaller final values of
\sig\ for each bin.  We then defined CO~WDs to be the 438 stars that
lie within $\pm$ 3\sig\ in any bin.  These stars are marked as large
black dots in Fig.~3.  Stars that lie between the CO~WDs and the best
He~WD candidates or that failed to meet the two additional criteria
described above were retained as marginal He~WD candidates.  These 17
stars are marked with small open circles in Fig.~3 and listed in the
lower portion of Table~1.

Having defined regions in the CMD occupied primarily by CO~WDs vs.\
He~WD candidates, we wish to assess the extent to which poorly
measured CO~WDs might be included among either the best 24 He~WD
candidates or the 17 marginal ones.  First we note that a close
examination of the \br\ vs.\ \R625\ CMD (Fig.~3, left panel) reveals 6
stars in the range \R625\ $= 22-24$ that are offset to the left of the
CO~WD sequence by about the same amount that the marginal candidates
are offset to the right of the sequence.  Thus it seems likely that
the marginal group of 17 stars includes a few poorly measured CO~WDs.
By contrast, the best 24 candidates have no such counterparts to the
left of the sequence and are likely to be a much cleaner set.

To more thoroughly assess the level of contamination among the He~WDs
we used artificial star tests in a manner similar to that described by
\citet{a08}.  We added artificial stars into the individual images,
and then reanalyzed the images using procedures identical to those
used to analyze the data originally (see \S 2).  A total of \about
100,000 artificial stars were introduced with magnitudes in the range
\R625\ = $20-28$ and colors matching the 0.53\msun\ cooling sequence
shown in the left panel of Fig.~3.  We then used the artificial star
results to construct 1000 simulated CO~WD sequences by picking stars
at random from the artificial stars that were recovered.  For each
simulated WD sequence, we extracted a number of stars equal to the
number of observed CO~WDs and with the same luminosity function, i.e.,
the same number of stars per 0.5 mag bin.  In each realization, we
counted how many of the artificial CO~WDs had landed in the regions of
the CMD previously defined to contain the 24 best and 17 marginal
He~WD candidates, respectively.

Among the 24 best candidates, we find that nearly all of the 18 stars
with \R625\ $<$ 23.5 are likely to be genuine He~WD candidates: the
average number of CO~WD interlopers in this region in the simulated
CMDs was only 0.8 $\pm$ 0.8.  Of the 6 stars with \R625\ =
$23.5-24.5$, about half may be interlopers, judging from the simulated
CMDs.  Thus roughly 20 of the 24 best candidates cannot be explained
as poorly measured stars; this accords with the results of a visual
inspection which revealed no sign that these stars were unusually
crowded in the images.  Among the 17 marginal candidates, the
simulated CMDs suggest that roughly half are likely to be poorly
measured CO~WDs.  We counted 2.8 $\pm$ 1.7 interlopers among those
with \R625\ $<$ 23.5 (vs.\ 9 observed) and 1.6 $\pm$ 1.2 interlopers
in the range \R625\ = $23.5-24.0$ (vs.\ 4 observed).  For \R625\ $>$
24.5, however, the artificial CMDs produced interlopers in numbers
comparable to the number of observed candidates.  We conclude that few
if any of the 4 faintest marginal candidates are likely to be genuine
He~WD candidates.

For the remainder of the paper we will focus primarily on the subset
of the 24 best He~WD candidates, the vast majority of which are
well-measured stars, while keeping in mind that perhaps half of the
brighter marginal candidates could also be genuine He~WD candidates.
Before going on to analyze their properties, however, we consider
other possible explanations for the presence of stars in the region of
the CMD that they occupy.  We note that there is little question that
they are cluster members: a uniform spatial distribution, as would be
expected for a group of foreground or background objects, is
definitively ruled out by their strong degree of concentration toward
the cluster center (see \S 6.1).  Our preliminary look at the proper
motions of these stars provides independent confirmation that nearly
all are cluster members; a full analysis of the proper motions is,
however, beyond the scope of the present paper.

We first consider the possibility that they may be cataclysmic
variables (CVs), eleven of which are now known in NGC~6397 (see
triangles in Fig.~3).  Given that two of these CVs land in the region
in the \br\ CMD occupied by the He~WD candidates, it is natural to ask
whether some of the He~WD candidates could instead be CVs; this
possibility was suggested by \cite{tb02} as a possible explanation for
the three candidates identified by \citet{tge01}.  One argument
against this possibility is that while all the known CVs were found as
counterparts to X-ray sources in Chandra imaging \citep{ghe01,cohn09},
none of the 41 He~WD candidates identified here have X-ray
counterparts down to the limit reached with existing Chandra imaging
(\lx\ \simlt\ 3 \x\ $10^{29}$~erg~s$^{-1}$) \citep{cohn09}.  Very few
if any CVs in the field have luminosities below this limit
(\citet{verbunt97}, \citet{pretorius07}, and references therein).

A second argument against the He~WD candidates being CVs comes from
comparing the locations of known CVs with the He~WD candidates in the
\hr\ vs.\ \R625\ diagram (Fig.~3, right panel).  Nine of the 11 CVs
have \hr\ colors significantly to the left of the main sequence.  This
implies excess emission through the \ha\ filter, and in turn the
likely presence of an \ha\ emission line (see \citet{gcc95} and
\citet{egc99} for the efficacy of this method in picking out
emission-line stars).  Even the two CVs that lie to the right of the
MS in \hr\ are still significantly \ha-bright compared to the WDs;
comparison to the white dwarf sequence is appropriate for most of the
fainter CVs, since their blue \br\ colors imply that their broad-band
spectra are dominated by the hot WD.  Thus all eleven of the CVs show
strong evidence of having \ha\ emission lines, which is a key
signature of accretion and nearly ubiquitous in CVs (e.g.,
\citet{williams83}).  By contrast, nearly all of the He~WD candidates
have \hr\ colors consistent with strong \ha\ absorption lines, and
very similar to those of the CO~WDs in the cluster.  Only one (ID=14)
has an \hr\ color close to any of the known CVs.

A second possibility to consider is that the He~WD candidates could
instead be detached WD-MS binaries.  Like CVs, such binaries would
naturally occupy the region between the MS and WD sequence
\citep{lopezAAS07}.  To explain the proximity of the candidates to the
CO~WD sequence in the \br\ vs.\ \R625\ CMD, however, would require
that the MS companions all have extremely low masses.  Combining the
observed CO~WD sequence with model MS stars from \citet{dcj08}, we
find that all but the brightest 4 He~WD candidates could be explained
as CO~WD/MS-star binaries only if the MS star had \mass\ \simlt\
0.15\msun.  \citet{cgc98} set a similar limit on the possible presence
of MS companions for the three original bright candidates using
multi-band WFPC2 data.  That the \hr\ colors of the He~WD candidates
closely resemble those of the CO~WDs also implies that if any of the
candidates were CO~WDs then the MS companions must contribute a small
fraction of the total light of the system.  As there is no reason to
think that binaries in NGC~6397 should favor such low-mass
companions---and exchange interactions certainly would not---we
conclude that it is unlikely that many of these stars are detached
WD/MS binaries.


\section{Hydrogen vs.\ Helium Atmospheres} \label{section:DAsvsDBs}


Use of the \ha\ filter enables us to determine whether hydrogen is
present or absent in the atmospheres of both the CO and He~WDs we
observe.  In Fig.~3 we have plotted a DB (helium atmosphere) cooling
sequence (thick solid line) along with the DA (hydrogen atmosphere)
cooling sequence (thick dash line).  We interpolated these tracks for
an assumed mass of 0.53\msun\ from the \citet{bwb95} CO~WD models.  In
the right panel, the same two tracks are also shown shifted 0.07
magnitudes to the right (thin lines) for reasons that will be
explained below.

The DA tracks span the entire range of observed magnitudes, while the
DB tracks begin at $T_{eff}=30,000$K (just below the so-called DB
gap---\citet{lwh86}).  In \br\ (left panel), the DA vs.\ DB tracks
separate by at most \about 0.1 magnitude in color (near \R625\ \about\
24)---too small a separation to definitively distinguish DA from DB
WDs using the present measurements.  In \hr\ (right panel), the tracks
are separated by \simgt 0.2 magnitudes in the range \R625\ = $22.5-25$
and as much as \about 0.4 magnitudes near \R625\ = 24.  Thus DA vs.\
DB WDs are more readily distinguished using this filter pair.  Note
that the DB tracks should be taken in this context to represent any
non-DA-type WDs, i.e., any WD with no \ha\ line, rather than DBs
exclusively.  We also note that the location of the DB track to the
right of the main sequence in the right panel of Fig.~3 is the result
of these stars being significantly hotter than the stars on the lower
main sequence.  Because the \ha\ filter is centered at $\lambda\
\simeq\ 6584$ \AA\ (i.e., at a longer wavelength than the \R625\
filter which is centered at $\lambda\ \simeq\ 6310$ \AA), hotter stars
will naturally produce lower ratios of \ha\ to \R625\ flux and
correspondingly larger values of \hr\ even in the absence of any
lines.

To determine which of the WDs in NGC~6397 are likely to be DA vs.\
non-DA WDs, we focused initially on the CO~WDs (large solid dots in
Fig.~3) and proceeded as follows.  First we considered only stars
brighter than \R625\ $=$ 24.5; for these stars, the separation between
the tracks in the \hr\ vs.\ \R625\ diagram is large enough relative to
the spread in the measured colors of the CO~WDs that we can
distinguish DA from non-DA WDs (Fig.~3, right panel).  An inspection
of this diagram reveals that nearly all of the CO~WDs follow the DA
cooling track, while there is no sign of any concentration of CO~WDs
around the DB track.  This strongly suggests that the vast majority of
CO~WDs in NGC~6397 are of the DA type.

We note that while the shape of the observed CO~WD sequence in the
\hr\ vs.\ \R625\ CMD is very similar to the shape of the theoretical
CO~WD track (thick dashed line), there is an offset between the color
of the track and the mean color of the CO~WDs.  We estimate that the
offset is \about 0.07 magnitudes, and surmise that it may be the
result of calibration uncertainties; shifted DB and DA tracks are
shown as thin solid and dashed lines, respectively.  The existence of
this small offset does not affect our conclusions below regarding the
relative abundance of DA vs.\ non-DA WDs in the cluster.  Nor does it
affect our measurement of the distance to the cluster (see \S 5) which
relies only on the \br\ vs.\ \R625\ diagram.

We can make a quantitative comparison between WDs in NGC~6397 and
those in the field by focusing on a portion of the CMD for which the
DA to non-DA ratio in the field is known.  \citet{tb08} find that in
the range $T_{eff} = 10,000-14,000K$ the ratio of DA to non-DA WDs is
about 4:1.  For the models we are using (and at the distance we
find---see \S 5), this temperature range corresponds to a magnitude
range of \R625\ $\simeq\ 23.5-24.5$.  In this magnitude range we
observe 126 CO~WDs, all but one of which lies closer to the DA track
than to the DB track; this number increases to four if we compare
instead to the shifted tracks.  If the DA/non-DA ratio matched that in
the field we should have seen \about 25 non-DA WDs.  The probability
of finding 4 when 25 are expected is 2.3\x 10$^{-7}$.  Thus it is
clear that the ratio of DA to non-DA WDs is considerably higher in
NGC~6397 than in the field.  Given the likelihood of a few poor
measurements among the 126 CO~WDs in the interval in question, it is
possible that all the observed CO~WDs in NGC~6397 have hydrogen
atmospheres.

This conclusion does not depend on our assumption that the CO~WD mass
is 0.53\msun, as the position of the tracks in the \hr\ vs.\ \R625\
diagram depends only very weakly on mass.  For example, the 0.51\msun\
cooling track lies much less than 0.01 magnitudes to the right of the
0.53\msun\ track in \hr.  Because the tracks are nearly vertical in
\hr\ vs.\ \R625, our conclusion is also largely independent of the
adopted distance.  Changing the distance modulus by up to 0.2
magnitudes would change the \hr\ colors of the tracks by less than
0.01 magnitudes.

These results demonstrate that the incidence of non-DA~WDs is much
lower in at least one globular cluster than in the field.  That more
than 96\% of the 126 CO~WDs we observe in NGC~6397 appear to have
hydrogen atmospheres confirms and extends the findings of
\citet{moehler04}.  \citet{krh05} found a similar preponderance of
DA~WDs in open clusters, and suggested that it could be the result of
the higher masses of CO~WDs in open clusters vs.\ the field.  That the
same phenomenon has now been seen to occur in a globular cluster,
whose CO~WDs have masses quite comparable to those in the field, makes
it unlikely that mass could be the primary factor in determining the
relative numbers of DA vs.\ non-DA stars.

Our findings also illustrate the power of \ha\ imaging for efficiently
classifying large numbers of WDs.  This is particularly valuable in
globular clusters in which WDs are faint and crowded by much brighter
stars, making spectroscopic observations very challenging.  In a
single study we have increased the number of WD classifications in
globular clusters by more than an order of magnitude.  \ha\ imaging
has been validated in its ability to pick out emission-line stars
(e.g., CVs) through follow-up spectroscopic observations
\citep{gcc95}.  Further testing of the method, in the context of
absorption-line stars, would be valuable to gain greater confidence
that it reliably distinguishes DA from non-DA WDs.

Finally, we note that the He~WDs also follow the DA cooling track (see
Fig.~3, right panel), showing that they too have hydrogen atmospheres.
Independent of any comparison to theoretical tracks, their
distribution is well-matched to those of the CO~WDs in \hr, and like
them shows no tendency to follow the DB track.  The only apparent
difference is that the spread in \hr\ colors for the He~WD candidates
seems to be somewhat larger than that for the CO~WDs.  This could be
due in part to their larger range in temperatures and masses at a
given \R625\ magnitude relative to CO~WDs.  We have explored the
effect of these parameters, however, and find that they are unlikely
to be able to fully explain the apparently larger spread; it is
unclear at present what the cause might be, if in fact it is
significant.


\section{The Distance to NGC 6397} \label{section:distance}


The distance to NGC 6397 has been subject to considerable uncertainty,
with reported values ranging from 2.2 kpc \citep{atts92} to 2.8 kpc
\citep{reid98}.  Most recently, \citet{hansen07} have detected the
bottom of the CO~WD sequence in an off-center field in NGC~6397 and
used it to determine a distance of 2.54~$\pm$~0.07~kpc (converting
from the quoted distance modulus) and an age of 11.47$\pm$0.47 Gyr.
Here we use the \citet{bwb95} theoretical WD cooling models
to determine a distance by matching the models to the
observed WD sequence in Fig.~3.  We assume a WD mass in the range 0.53
$\pm$ 0.02 \msun, as suggested by \citet{renzini96} based on the
luminosities of the termination of the asymptotic giant branch (AGB),
the RGB tip, the horizontal branch, and post-AGB stars.  This mass
range is also in accord with spectroscopic mass measurements of CO~WDs
in NGC~6397 \citep{moehler04}, and encompasses the mass of 0.51\msun\
determined by \citet{hansen07} (see their Fig.~19).  In view of the
good match between the DA cooling sequence and the location of the
CO~WDs in the \hr\ CMD (see \S 4), we also assume that all the CO~WDs
are of the DA type.

Starting with a 0.53\msun\ DA WD model in the \br\ diagram, we tested
distances from $2.20-2.80$~kpc at 0.01~kpc intervals and found the
best fit value as follows.  First we determined extinction values in
both filters assuming $E(B-V) = 0.18$.  For stars with colors of
CO~WDs in NGC~6397, and adopting the reddening law of
\citet{seaton79}, we obtained $A_{435} = 0.75$ and $A_{625} = 0.48$;
these values are very weakly dependent on spectral type.
We then shifted the theoretical cooling sequence to account for the
distance and for the extinction in both filters.  Next we selected all
stars within $\pm$0.8 magnitudes in \br\ color of the theoretical
sequence and in the range \R625\ $= 18.9-26.0$, excluding known CVs.
We then computed the offset in \br\ color between each of these stars
and the theoretical sequence at the same \R625\ magnitude.  To select
stars likely to be CO~WDs, we computed the standard deviation of these
color offsets (using the bins described in \S 3), iterating several
times and removing stars with $>$~2\sig\ deviations at the end of each
iteration.  This fairly stringent requirement was imposed in order to
reliably remove non-CO~WDs from consideration.  The best fit distance
was taken to be the distance that yielded a mean \br\ color offset
closest to zero.  The best fit was achieved for a distance of
2.34~kpc, or an apparent distance modulus of $(m-M)_{625} = 12.33$;
the result is shown in both panels of Fig.~3 (long-dash lines).

The main contributions to the uncertainty in the distance come from
uncertainties in the CO~WD mass, the reddening, and the photometric
calibration.  Examining the WD models, we find that a change of $\pm$
0.02\msun\ from a CO~WD mass of 0.53\msun\ corresponds to a change of
\about 0.05 in the \R625\ magnitude for the range of WD temperatures
in question.  This translates to a contribution to the uncertainty in
the distance of \about 0.05~kpc.  The uncertainty in $E(B-V)$ is 0.02
magnitudes \citep{harris96}, which translates to a uncertainties of
0.053 in $A_{625}$ and 0.03 in $A_{435}-A_{625}$.  Rerunning the
model-fitting algorithm described above using $E(B-V) = 0.18 \pm\
0.02$, we find that the uncertainty in the reddening results in an
uncertainty in the distance of \about 0.05 kpc.  Finally, a maximum
uncertainty in the photometric calibration of up to 0.1 mag would
correspond to a distance uncertainty of \about 0.11~kpc.

Combining these contributions to the uncertainty in quadrature yields
a final estimate for the uncertainty in the distance of 0.13~kpc.
Thus, with an assumed CO~WD mass of 0.53\msun, we find a distance of
2.34~$\pm$~0.13~kpc.  We note that our final result is in reasonable
agreement with the distance determined by \citet{hansen07} if we adopt
their value of 0.51\msun\ for the mass of all but the very faintest
CO~WDs (see their Fig.~19).  In this case our distance increases to
2.39~$\pm$~0.13~kpc, which is to be compared with their value of
2.54~$\pm$~0.07~kpc.


\section{Discussion} \label{section:discussion}


The sequence of 24 helium-core WD candidates identified here
quadruples the number known in NGC~6397.  This is the first such
extensive sequence of individually identified He~WDs found in a
globular cluster.  Their presence illustrates the complexity of WD
sequences in GCs, which has been extensively explored by
\citet{hurley03}.  A large population of He~WDs has also been proposed
to explain the total numbers and range of colors of WDs observed in
$\omega$ Cen \citep{calamida08}.  Our ability to identify individual
He~WD candidates in NGC~6397 presents a unique opportunity to study a
large group of these rare stellar remnants all at the same, well-known
distance.  Given their connection to binary stars and/or stellar
interactions, they provide a fresh window into the inner workings of a
globular cluster.

\subsection{Photometric vs.\ dynamical masses of the He~WDs}

We begin by examining the location of the He~WD candidates in the \br\
vs.\ \R625\ CMD, comparing their magnitudes and colors to theoretical
He~WD models. These He~WD models have been computed with the same code
and input physics as described by \citet{sar02}, and their synthetic 
spectra according to \citet{rsa02}. Fig.~4 shows model
sequences for masses ranging from 0.175-0.45\msun\ and Z=0.0002
progenitors, a metallicity appropriate for NGC~6397
\citep{harris96}. Because magnitudes (and to a much lesser extent
colors) depend somewhat on the thickness of the remnant's hydrogen
layer (and cooling age much more so---see below), we have computed two
sets of models characterized by different H-layer masses.  Models with
``thick'' envelopes (left panel) have H-layer masses ranging from
$5\times10^{-3}$\msun\ for the least massive model down to
$4\times10^{-4}$\msun\ for the 0.45\msun\ WD model. These models were
obtained by assuming that binary evolution leads to stable mass
transfer; computational details can be found in \citet{sar02}. In this
case mass loss from the WD progenitor ends when the envelope mass
becomes small enough that an extended structure cannot be supported
any longer and then shrinks within the Roche-lobe.  Models thus
computed represent an upper limit to the possible H-layer thickness
that a He-WD model of given mass can have. In these models, nuclear
burning (mostly through the p-p chain) is a main source of energy along
the WD cooling phase.  On the other hand, ``thin'' envelope models
(Fig.~4 right panel) were obtained by modifying the envelope structure
of pre-WD models in such a way as to render hydrogen nuclear burning
at the base of the envelopes negligible, while keeping the total WD
mass fixed.  In our models, this translates into ``thin'' H-layers
with masses between $1\times10^{-4}$\msun\ and
$5\times10^{-5}$\msun. Our sets of ``thick'' and ``thin'' models
represent the two extremes in terms of cooling timescales for He-WD
stars. They also bracket reality in terms of mass-radius relations for
these stars. Even thinner H-layers than those adopted in our models
would not translate into noticeable smaller radii (fainter magnitudes).

Comparison to either set of models shows that most of the 24 He~WD
candidates (large circles) have magnitudes and colors consistent with
masses in the range \mass\ \about 0.2-0.3\msun.  This is in good
agreement with the mass of \about 0.25\msun\ determined for the one
candidate for which a spectrum has been obtained (ID=2 in Table 1;
\citet{egc99}).  Comparing the two panels, we see that with thinner
envelopes (right panel) the inferred masses shift to slightly lower
values on average.  This effect is particularly pronounced at the
bright end (see also \citet{hkr03}), such that the brightest candidate
lands outside the region spanned by the thin-envelope models in the
right panel.  Within the measurement uncertainties, the remaining
candidates are all within the region spanned by either set of models.
This includes candidates 4, 5, and 6, first identified by
\citet{tge01}, whose \vi\ colors measured from WFPC2 images appeared
noticeably redder than 0.2\msun\ models \citep{sar02}.  We speculate
that the discrepancy may be the result of a bias toward redder
measured colors for stars near the detection limit in the WFPC2
observations.  In both panels of Fig.~3 the marginal candidates (small
open circles) lie along tracks predicted for 0.35-0.45\msun\ He~WDs.

An independent measure of the average mass of the He~WD candidates as
a group can be obtained by examining their radial distribution within
the cluster.  The dynamical relaxation time at the half-mass radius
(\rh\ = 2\minspt 33) is \about 3 \x\ $10^8$ years; at the center it is
less than $10^5$ years \citep{harris96}.  As all the He~WD candidates
are well inside the half-mass radius (and half of the best 24 are less
than 30\asec\ from the center), their radial distribution should
reflect their masses.  Indeed, a high degree of mass segregation in
the central regions of NGC~6397 has been well documented
\citep{ksc95}.

In Fig.~5 we plot the cumulative radial distribution of the 24 best
He~WD candidates (solid black line) along with that of several other
populations in the cluster: CO~WDs (short-dash line), turnoff stars
(long-dash), blue stragglers (dash-dot), and the 17 marginal He~WD
candidates (solid grey line).  We also plot the cumulative
distribution that would be expected for a spatially uniform
distribution of objects (dotted line).  Applying the
Kolmogorov-Smirnov test to compare distributions, we find first of all
that the 24 best He~WDs have a 2\x $10^{-7}$\% chance of being drawn
from a spatially uniform distribution, which definitively rules out a
population of foreground or background interlopers.  They are also
significantly more centrally concentrated than the CO~WDs (0.02\%
probability of being drawn from the same population) and main-sequence
turnoff stars (0.2\% probability).  This strongly suggests that, on
average, their masses are in excess of the turnoff mass of \about
0.8\msun.  These comparisons do not take account of incompleteness,
except that we excluded CO~WDs fainter than \R625\ = 24.5 (the
magnitude of faintest of the good He~WD candidates).  The greater
difficulty of detecting He~WDs as compared to turnoff stars in the
central regions would only make the difference in radial distribution
more pronounced.  Of all the groups to which we compare the He~WDs,
they most closely resemble the blue stragglers (44.1\% probability),
which have a range of masses \about 0.8-2\msun\ \citep{saffer02,
demarco05}.

In contrast to the 24 best He~WD candidates, the 17 marginal
candidates are not nearly as concentrated to the center as the blue
stragglers (K-S probability = 0.16\% of being drawn from the same
population; see Fig.~5).  They most closely resemble the CO~WDs
(94.3\%), though are also compatible with turnoff stars (48.8\%).  The
probability that these 17 stars are drawn from the same population as
the 24 best candidates is only 1.0\%.  Given the significant degree of
contamination that we expect among these marginal candidates (roughly
half are likely to be poorly-measured CO~WDs---see \S 3), the
similarity of their radial distribution to that of the CO~WDs is
perhaps not surprising.  Still, if the other half really were He~WDs,
and these had masses comparable to those of the 24 best candidates,
one might expect a greater degree of central concentration.  We
conclude that either the degree of contamination is greater than we
estimated or that the He~WDs that do exist among these marginal
candidates may not have the large dynamical masses characteristic of
the 24 best candidates.  In view of the uncertain status of the
marginal candidates, we hereafter focus our attention primarily on the
24 best candidates.

\subsection{Nature of the binary companions}

The mismatch between photometric masses determined from the CMD and
dynamical masses determined from the radial distribution for the 24
best He~WD candidates can be understood if they have binary companions
of sufficient mass.  Such companions are expected on evolutionary
grounds (see \S 1) and earlier observations found evidence for them
\citep{cgc98, egc99, tge01}.  Here we use the measured dynamical vs.\
photometric masses of the He~WDs to place new constraints on the
nature of these companions.  

Given that the bulk of the He~WD candidates have photometric masses in
the range 0.2-0.3\msun, whereas the dynamical masses resemble those of
the much more massive blue stragglers ($0.8-2$\msun), the companions
must make up this large mass deficit.  We note that the blue straggler
mass range found by De~Marco et al. (2005) is based on six of the
brightest blue stragglers in NGC 6397 and thus may not be
representative of the entire population of 24 blue stragglers in our
sample.  To estimate the typical blue straggler mass, we have
performed cored power-law fits to the radial distributions of turnoff
stars and blue stragglers \citep{cohn09}.  Taking these two groups to
be in approximate thermal equilibrium results in a blue straggler mass
of 1.2~$\pm$~0.1\msun.  Adopting this as the typical mass for the
He~WD binary systems suggests a companion mass of \about 1\msun.  This
rules out main-sequence companions, as MS stars near the turnoff mass
would dominate the light and cause the He~WD binaries to be
significantly brighter and redder than is observed.  CO~WDs with
masses \about 0.53\msun (like those currently being produced in the
cluster) are also ruled out, since they fall well short of the
required companion mass.  We have also compared the observed radial
distribution of the He~WDs to the expected distribution of 1.6\msun\
stars (i.e., the dynamical mass of a He~WD binary containing a neutron
star) and find that 1.4\msun\ companions are ruled out, even when we
allow for incompleteness in the He~WD distribution.  Thus we conclude
that the required typical mass of the unseen companions of about
1\msun\ is most compatible with heavy WDs.

The non-detection in X-rays of any of the 41 He~WD candidates further
supports the view that the companions are white dwarfs rather than
neutron stars.  NSs should have been spun up to become MSPs during the
mass-transfer stage.  For comparison, the 19 MSPs in 47~Tuc have 
\lx\ \about $10^{30}-10^{31}$ erg s$^{-1}$ \citep{bgh06}; if any of the
He~WDs in NGC~6397 had comparable MSP companions they should have been
seen in existing Chandra imaging that reaches 
\lx\ \simlt\ 3 \x\ $10^{29}$ erg s$^{-1}$ \citep{cohn09}.

Our finding that the companions are most likely to be white dwarfs is
in good agreement with the conclusions of \citet{hkr03}, who argue on
theoretical grounds that the companions to the He~WDs in NGC~6397
should be WDs and not neutron stars (NSs).  A key part of their
argument is that mass transfer from a \about 0.8\msun\ turnoff star
onto a NS would be stable, whereas onto a WD it would be unstable.
Stable mass transfer would yield relatively wide binaries which would
be vulnerable to rapid breakup in the dense environs of the center of
NGC~6397.  Unstable mass transfer onto a WD would instead yield tight
binaries (following a period of common-envelope evolution) which would
be more resilient.  We note that the latter scenario would apply to
mass transfer onto white dwarfs with mass less than \about 1\msun, but
not for more massive WDs---a point we will return to below.

\subsection{Formation rate of He~WDs and implications for binary population}

We can use cooling ages from the He~WD models to consider what the
distribution of He~WDs in the CMD implies about the formation rate of
these stars in NGC~6397.  This in turn can provide insight into the
viability of different models and/or formation scenarios.  The
analysis is complicated by the fact that cooling ages are highly
dependent upon the assumed mass of the residual hydrogen layer, with
thicker layers leading to slower rates of cooling.  Because thick
hydrogen layers support hydrogen burning, WD evolution along the
cooling sequence for thick-envelope models is completely dominated by
nuclear burning on the very long timescales dictated by the pp-cycle.
By contrast, thin H envelopes render nuclear burning negligible so
that for thin-envelope models, WD evolution is driven solely at the
expense of the star's internal heat.  The differing rates of cooling
can be seen in Fig.~4 by comparing the left and right panels, where
``+'' symbols are used to indicate cooling ages.  In the left panel,
marked ages are $1-13$ Gyr in 1-Gyr intervals.  In the right panel,
they indicate ages of 0.001, 0.0025, 0.005, 0.0075, 0.01, 0.025, 0.05,
0.075, 0.1, 0.25, 0.5, 0.75, 1.0, 1.25, 1.5, 1.75, 2.0, 2.25, 2.5, and
2.75 Gyr.  Thus, for example, the brightest of the three He~WD
candidates identified by \citet{tge01} at \R625\ $\simeq$ 22.5 has a
cooling age of \about~8~Gyr according to the thick-envelope models
(left panel) vs.\ an age of only \about~0.1-0.25~Gyr according to the
thin-envelope models (right panel).  We caution that the very youngest
ages shown for the thin-envelope models (on the order of 1~Myr) are
uncertain by a factor of a few due to the fast initial rate of
evolution and the somewhat ad-hoc nature of the models.

Here we examine the implications of thin vs.\ thick envelopes in turn.
It should be emphasized that these models represent only two possible
sets of H~envelope masses.  However, they
represent the two opposite extremes both in terms of
cooling ages and mass-radius relations and thus should bracket
reality.  

Considering the thin-envelope models first (Fig.~4, right panel), we
see that the 24 best He~WD candidates have implied cooling ages
ranging from \simlt 1~Myr for the brightest candidate to about 1~Gyr
for the faintest.  The implied formation rate is then \about
24~Gyr$^{-1}$.  This is higher than the rate at which RGB stars and
compact objects collide in the central regions of NGC~6397.  Assuming
then that the He~WDs are instead the products of pre-existing
binaries, the formation rate is determined by the fraction of stars
that have a close white dwarf companion when they ascend the giant
branch.  The implied binary fraction can thus be determined by
comparing the rate of formation of He~WDs to the rate at which stars
are turning off the main sequence in the same region.  Counting stars
above the turnoff at \R625\ = 16.1, we find \about 700 subgiant and
red giant stars.  Considering the \about 1.5~Gyr spent in these stages
combined \citep{psh98}, this implies that stars are turning off the
main sequence at a rate of about 470 Gyr$^{-1}$ at the present epoch.
Similar rates are implied by the number of horizontal-branch stars
observed (48, which given a lifetime of \about 0.1~Gyr implies a rate
of 480~Gyr$^{-1}$) and by the number of CO~WDs with ages of 1~Gyr or
less (\about 480 after correcting for incompleteness).  Thus the
implied binary fraction is 24/470 or about 5\%.  This is a small
enough fraction that it is not surprising that the horizontal-branch
and CO~WD populations do not appear to be depleted by losses to He~WD
formation.

The \about 5\% binary fraction required to account for the observed
He~WDs in the context of thin-envelope models represents a lower limit
on the total binary population in the central regions of the cluster,
as it accounts only for the particular class of binary that yields
He~WDs.  The \simlt 5-7\% binary fraction found among MS-MS binaries
in the central region of the cluster by \citet{cb02} would be in
addition to the He~WD progenitor binaries.  The total will be less
than the sum if some MS-MS binaries are themselves progenitors of the
binaries that lead to the He~WDs, i.e., if a CO~WD replaces one of the
MS stars through an exchange collision before the other MS star turns
off the main sequence (or at least before it climbs far enough up the
giant branch to fill its Roche lobe).  Such exchanges are not uncommon
in N-body simulations of high-density clusters (e.g.,
\citet{ivanova06}).

Interpreting the He~WDs in the context of thin H-envelope models
implies a binary fraction that is quite plausible.  However, it does
present one problem.  The implied cooling ages of the brightest four
candidates are very young, in the range of only \about 1-2~Myr.  It
would be surprising to catch any He~WDs so early in their evolution,
even considering the factor of a few uncertainty in the cooling ages
of such young objects.  These stars cannot just be a small number of
interlopers unrelated to the cluster.  Three have already been shown
to be proper-motion members \citep{cb02}, and our preliminary analysis
shows that the fourth is as well.  Such young ages would imply an
implausibly high rate of formation in the very recent past, nominally
on the order of 2~Myr$^{-1}$.  Even taking the uncertainty in their
ages into account, the present-day rate of formation would be more
than an order of magnitude greater than the average rate over the past
1~Gyr.  The current formation rate would also have to be higher than
the total rate at which stars are turning off the main sequence at
present (\about 0.5~Myr$^{-1}$).  Yet only a fraction of these turnoff
stars can be members of binaries that will produce He~WDs.  Thus it
seems highly unlikely that these bright stars can be explained as
He~WDs with thin H envelopes.

Examining the cooling ages in the context of thick H-envelope models
instead (Fig.~4, left panel), the He~WDs would have ages ranging from
\simlt 1~Gyr to \about\ 13~Gyr.  In this case, the formation rate
problem for the brightest candidates disappears, owing to the much
slower rate of cooling associated with thick envelopes.  The implied
rate of formation in the recent past is \about 5~Gyr$^{-1}$, easily
compatible with a binary formation scenario, as it would require that
only \about 1\% of the stars turning off the main sequence produce
He~WDs in the present epoch.  This rate of formation is also low
enough that collisions between RGB stars and CO~WDs could make a
significant contribution to their production.  The rate is also not
too different from the long-term formation rate of \about 2~Gyr$^{-1}$
implied by seeing 24 He~WD candidates with ages up to \about 13~Gyr.
Indeed, a moderate increase in production rate would not be surprising
considering the inevitable dynamical evolution of the cluster on this
timescale.

In summary, either thick or thin envelopes can explain the
observations with reasonable implications for the formation rate of
He~WDs (\about $2-24$ Gyr$^{-1}$) and the binary fraction in the
cluster (\about $1-5$\%).  The one exception is that the brightest
candidates cannot be readily explained in the context of thin
envelopes, as they would require an exceedingly high rate of formation
in the very recent past.

Additional insight into the population of He~WD progenitor binaries
can be gleaned by examining the distribution of He~WD masses, as
inferred from their positions in the CMD.  The mass of the degenerate
core that will be left behind when a giant loses its envelope to a
binary companion depends on how far up the RGB the giant has climbed
before overflowing its Roche lobe.  This in turn depends on the
orbital period of the binary: longer period systems, being wider,
allow the giant (and the mass of its He core) to grow larger before
the envelope is lost.  The He~WD masses we deduce from the CMDs are
mostly rather small.  All but 2 of the 24 best candidates have implied
masses in the range \about $0.2-0.3$\msun.  This in turn implies that
the progenitor binaries have periods in the range \about $1-20$ days
\citep{rpj95}.  Longer-period binaries containing CO~WDs and MS
turnoff stars or giants appear to be significantly more rare, given
our finding (see \S 3) that at most \about 6 of the marginal
candidates with \R625 = $22-24$ are likely to be genuine He~WDs.

\subsection{Termination of the He~WD sequence}

Irrespective of the thickness of H envelopes, the He~WD sequence comes
to a relatively abrupt end near \R625 \about 24.  The dearth of He~WD
candidates at fainter magnitudes is readily visible in the Fig.~4 CMDs
(recall from \S 3 that our artificial star tests show that the four
faintest marginal candidates are likely to be poorly measured CO~WDs).
By contrast, the number of observed CO~WDs is continuing to rise below
\R625\ \about 24.  The lack of fainter He~WDs cannot be the result of
incomplete recovery of faint stars.  Artificial star tests show that
in the bins from \R625\ = $24.5-25.0$ and $25.0-25.5$, the
completeness is still \about 60\% and 50\%, respectively (see Fig.~6,
top panel).  Thus it is unlikely that any significant number of
fainter He~WDs is present.  When we correct for incompleteness (see
Fig.~6, panels 2-4), the CO~WD luminosity function continues to rise
to the limit of our observations, whereas the number of He~WDs is
clearly in decline.

To further examine the significance of the absence of convincing He~WD
candidates below \R625 \about\ 24.5, we measured the offsets in color
from the 0.53 $M_\odot$ CO~WD cooling track for each of the CO~WD and
He~WD candidates.  Fig.~7 shows color offset distributions for stars
with \R625\ = $19-24.5$ (left panel) and \R625\ = $24.5-26$ (right
panel).  The distribution of the brighter stars is clearly bimodal,
with the He~WD candidates forming a peak centered \about 0.4
magnitudes redward of the CO~WD sequence.  For the fainter stars, the
color spread is larger due to growing measurement uncertainties.
Still, there is no sign of any excess of stars on the right side
($\Delta$(\br) \simgt 0.3) where He~WDs would be expected to lie.  We
estimate that at most \about 10 He~WDs (2\sig\ upper limit) could hide
in the wings of the CO~WD distribution in the magnitude range \R625 =
24.5-26.

The significance of the termination of the He~WD sequence is different
in the context of thin vs.\ thick H-envelope models.  If the hydrogen
envelopes are thin, the sequence ends at \about 1~Gyr, well below the
age of the cluster.  Such a termination was predicted by
\citet{hkr03}, who suggested that within \about 1~Gyr, binaries
containing He~WDs would either merge via gravitational-wave radiation
or be broken up in exchange interactions.  The He~WDs, as the
lowest-mass stars in most such exchanges, would typically be removed
from the binary and quickly redistribute themselves according to their
single-star mass.  While we find that the sequence extends about one
magnitude fainter than predicted by \citet{hkr03} (who suggested that
none fainter than those identified by \citet{tge01} would be found in
deeper searches), our finding agrees with their prediction within the
large uncertainties in cooling age associated with the unknown
H-envelope thickness.  In particular, while their ``moderate'' H
envelope models put the cooling age of the faintest \citet{tge01}
candidate at \about 1~Gyr, the ``thin'' H-envelope models we use here
result in more rapid cooling, such that the He~WDs are a magnitude
fainter at 1~Gyr.

While our observations confirm the prediction by \citet{hkr03} that
the sequence should end at \about 1~Gyr, there is still a puzzle.
This prediction was made specifically for the cluster core, assuming a
central mass density of 1.5\x $10^6$ \msun\ pc$^{-3}$.  Yet few, if
any, of the He~WD candidates are actually in the cluster core, which
is very small.  \citet{sosin97} found a core radius of $1.5-8$\asec\
at 95\% confidence, while our preliminary analysis of the present data
set suggests a radius in the range \about $5-7$\asec.  Outside the
cluster core, in the so-called ``cusp'' region, the density falls off
approximately as (r/r$_c$)$^{-2}$.  Given that a binary's lifetime to
exchange is inversely proportional to the density, a binary that would
last only 1~Gyr in the core would survive more than a Hubble time at
just 4\rc.  Depending on the core radius, between half and all the
He~WD candidates are outside 4\rc, where the lifetime to exchange
would be even longer.  Thus it is not clear that the termination of
the sequence can be explained in this way.  N-body simulations that
can more directly assess the longevity of binaries, without the
simplifying assumption that they reside at a fixed radius, would be
valuable. 

In the context of thick envelopes, as illustrated in the left panel of
Fig.~4, the faintest of the He~WD candidates would have cooling ages
of \about 13~Gyr.  Very old He~WDs could have evolved from
main-sequence stars with masses of \simgt 2\msun, whose MS lifetimes
are \simlt 1~Gyr \citep{psh98}.  Such stars have core masses \simlt
0.3\msun\ as they are leaving the main sequence \citep{nelemans01a}.  Using
the present thick-envelope models, the total inferred ages would be
uncomfortably high considering recent age determinations for the
cluster: 12~$\pm$~0.8~Gyr \citep{att00} and 11.47~$\pm$~0.47~Gyr
\citep{hansen07}.  This alone should not rule out thick envelopes,
however.  Given the strong dependence of cooling rates on H envelope
mass, even slightly lower envelope masses would likely bring the ages
into agreement.  Moreover, the argument against thick envelopes put
forward by \citet{hkr03} rests in large part on their finding that
binaries in the core will either merge or be broken up after ~1 Gyr.
They also suggest that common-envelope (CE) evolution will lead to
thin H envelopes.  However, in view of the location of many of the
binaries outside the cluster core, and the lack of direct evidence
that CE evolution leaves behind thin envelopes, we conclude that thick
envelopes are not ruled out.

Another possibility is that the He~WDs could be a mixture of systems,
some with thick and others with thin H envelopes.  If common-envelope
evolution does lead to He~WDs with thin envelopes \citep{hkr03}, then
all systems for which a \about 0.8\msun\ giant overflows its Roche
lobe onto a CO~WD companion with \mass~\simlt~1\msun\ should have thin
envelopes and cool off quickly.  For giants with CO~WD companions of
larger mass, however, the mass transfer would be stable.  If thick
envelopes were the result, then this subset of systems would cool much
more slowly.  The brightest systems in NGC~6397 could potentially be
explained as such higher-mass binaries.  

This hypothesis is appealing in that it could help make sense of two
other ways in which the four brightest He~WDs seem to set themselves
apart from the remaining 20 He~WD candidates.  First, their apparently
greater concentration toward the center (median radial offset of
12\asec\ vs.\ 33\asec\ for the fainter 20) could be explained as the
result of mass segregation.  Second, the 1.6-magnitude gap between the
four brightest He~WDs and the remaining systems could be the result of
two very different rates of cooling.  Judging from the models shown in
Fig.~4, the youngest thick-envelope stars would have ages similar to
the oldest thin-envelope stars (\about 1~Gyr).  He~WDs formed via
stable mass transfer should have relatively long periods following the
mass transfer episode: \about $2-40$ days for $0.2-0.3$\msun\ He~WDs
\citep{rpj95}.  These binaries would be more vulnerable to exchange
interactions than post-common-envelope binaries which would have much
shorter periods.  Nevertheless, at the range of radii where the four
bright He~WDs lie (\about 6-24\asec), systems with 2-day periods could
survive \about 0.6-10 Gyr, while systems with 40-day periods could
survive \about 0.08-2 Gyr before the He~WD would be exchanged out of
the binary (cf. \citet{hkr03}).  We surmise that the few we see
concentrated to the cluster center could be the remnant of a larger
population of such systems.

Finally, we note that the possibility that some He~WDs could be torn
away from their binary companions via exchange interactions raises the
question of where these stars would be at present, and whether a
second population of He~WDs might be observable in the cluster.  It
may be that such stars would acquire high enough recoil velocities
during exchange interactions that they would simply be ejected from
the cluster.  Escape from the cluster would follow, for example, if a
0.2\msun\ He~WD in a 10-day binary with a 1\msun\ CO~WD acquired a
velocity comparable to its orbital velocity of \about 90 km s$^{-1}$.
If, however, He~WDs are retained in the cluster following exchange
interactions, their half-mass radius should be larger than that of the
cluster as a whole (\rh\ \about 140\asec), due to their very low
masses.

To estimate how many such isolated He~WDs would still lie within the
field of view spanned by the present observations, we first note that
any thin-envelope He~WDs older than 3~Gyr would be fainter than the
detection limit (see Fig.~4, right panel).  Scaling from the 24 He~WDs
with ages of 1~Gyr or less, we estimate that another \about 50 would
have cooling ages in the range $1-3$~Gyr.  Using the radial
distribution of 0.25\msun\ stars in the Fokker-Planck models of
NGC~6397 developed by \citet{dull96}, and assuming that all 50 of
these He~WDs have been exchanged out of binaries, we estimate that
\about 8 isolated He~WDs would currently lie within a radius of
1.5\amin.  This is just barely compatible with our finding above that
up to \about 10 He~WDs could hide in the broad wings of the CO~WD
distribution at very faint magnitudes in the CMD.

Using the same approach we can ask how many isolated He~WDs would be
present in the off-center field studied by \citet{richer08}.  Assuming
that He~WDs could be detected with ages up to \about 10~Gyr in that
field, we estimate that \about 7 isolated He~WDs would be present in
their full ACS/WFC field.  Taking a look at the CMD for that field
(see their Fig.~3), there is no clear sequence of He~WDs, but there
are nevertheless several stars in the region such stars would occupy.
Since it is possible that these stars are unrelated to the cluster, we
also inspected their proper-motion-cleaned CMD (their Fig.~5).  In
this CMD, only 2 stars are present in the He~WD region.  However, this
CMD excludes roughly 40\% of the area of the original field (the area
not covered by the 1st-epoch WFPC2 data), such that we would predict
the presence of \about 4 isolated He~WDs, consistent with what is
seen.  We conclude that current observations do not rule out the 
existence of a second population of very faint, isolated He~WDs in 
NGC~6397.


\section{Summary and Conclusions} \label{section:summary}

We have identified a set of 24 faint blue, \ha-faint stars in the
globular cluster NGC~6397 that are good candidates to be white dwarfs
with helium cores.  This is the first extended sequence of He~WDs
found in a globular cluster.  The masses of the He~WD candidates that
we infer from comparisons to the colors and magnitudes of theoretical
cooling models are in the range \about $0.2-0.3$\msun.  Dynamical
masses inferred from the radial distribution are significantly higher,
on the order of 1.2\msun, implying the presence of binary companions.
Constraints from the observed colors of the He~WDs and their lack of
X-ray counterparts suggest that the companions are most likely to be
heavy white dwarfs.

The rate of formation of He~WDs that can be inferred from their
distribution in the color-magnitude diagram is highly dependent on
their rate of cooling, which in turn depends on the uncertain
thickness of their hydrogen layers.  We compare the observed systems
to two set of models, one with thick and the other with thin H
envelopes, and infer that \about $1-5$\% of stars currently turning
off the main sequence eventually become He~WDs.  This can be
understood if \about $1-5$\% of main-sequence stars have or obtain
CO~WD companions by the time they begin evolving up the RGB.  Initial
binary periods of \about $1-10$ days are required to produce He~WDs
with the masses we observe.  We note that this binary fraction is a
lower limit to the overall binary fraction in the cluster, as it
accounts only for the subset of binaries that yields He~WDs with CO~WD
companions.

The He~WD sequence ends about 1.5 magnitudes above the limiting
magnitude of our study.  The interpretation of this termination is
different in the context of thick vs.\ thin H envelopes.  In the case
of thick envelopes, the faintest observed systems have ages comparable
to the age of the cluster.  In the case of thin envelopes, the
faintest have ages of \about 1 Gyr.  Neither the thick nor thin
envelopes considered here provide a completely satisfactory
explanation for the current observations.  It does seem clear that the
brightest four He~WDs have thick H envelopes, as thin envelopes would
imply very young ages that in turn would require an implausibly high
rate of He~WD formation at the present epoch.  If the rest have thin
envelopes then the termination of the He~WD sequence must be
explained.  The suggestion by \citet{hkr03} that mergers and/or
exchanges destroy the binaries after \about 1 Gyr appears viable, and
could perhaps be tested by wider-area searches for faint He~WDs in
off-center fields.  If all He~WDs have thick H~envelopes, then the
envelope thickness must be somewhat lower than what we have assumed
here in order for the oldest observed He~WDs to have cooling ages
compatible with the age of the cluster.

Nearly all of the CO~WDs brighter than \R625\ \about\ 24.5 ($T_{eff} >
10,000$K) have \hr\ colors that strongly suggest that they have the
broad \ha\ absorption lines characteristic of DA-type
(hydrogen-atmosphere) WDs.  This suggests that whatever mechanism is
operating to produce DBs (or other non-DA type WDs) in the field is
ineffective in NGC~6397.  White dwarf mass is unlikely to be the
primary factor determining the relative abundance of DA vs.\ non-DA
WDs since the masses of CO~WDs in NGC~6397 are similar to those of
CO~WDs in the field.  The 126 stars in our sample of tentatively
classified CO~WDs represents the largest sample of WDs for which
atmospheric classifications have been obtained in a single cluster.
While further validation of the method is needed to ensure that it
reliably distinguishes DA from non-DA WDs, \ha\ imaging should be a
powerful tool to classify white dwarfs in star clusters.

Finally, we have compared the observed CO~WD sequence to theoretical
cooling models to measure a distance to NGC~6397.  With an assumed
CO~WD mass of 0.53\msun\ and reddening of $E(B-V)$ = 0.18, we obtain a
distance of 2.34~$\pm$~0.13~kpc.

\vskip 0.5in
 
We gratefully acknowledge Pierre Bergeron for providing CO~WD models
in HST/ACS filters, and for thoughtful comments; Onno Pols and Aaron
Dotter for making useful RGB and MS models available online; Liliana
Lopez and Jason Kalirai for helpful and interesting discussions; and
the anonymous referee for providing helpful comments which improved
the paper.


\clearpage
\begin{figure}[h!] \begin{center} 
\includegraphics[scale=0.80]{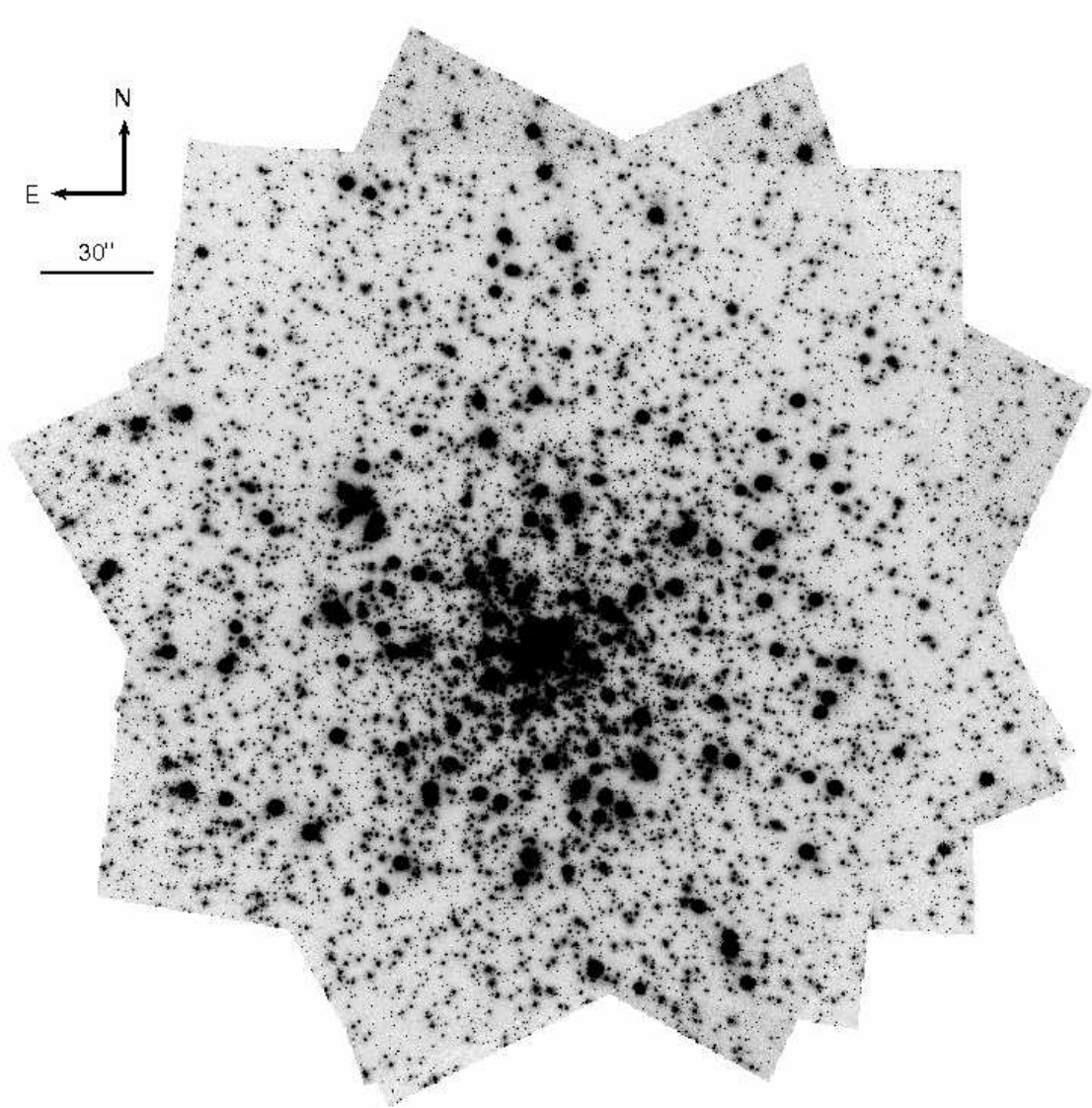}
\caption[]
{Mosaic image of NGC~6397 in the \b\ filter, constructed from images taken on
five visits spread over one year.}
\label{fig:cluster_image} 
\end{center} \end{figure}

\clearpage
\begin{figure}[h!] \begin{center} 
\includegraphics[angle=90, scale=0.85]{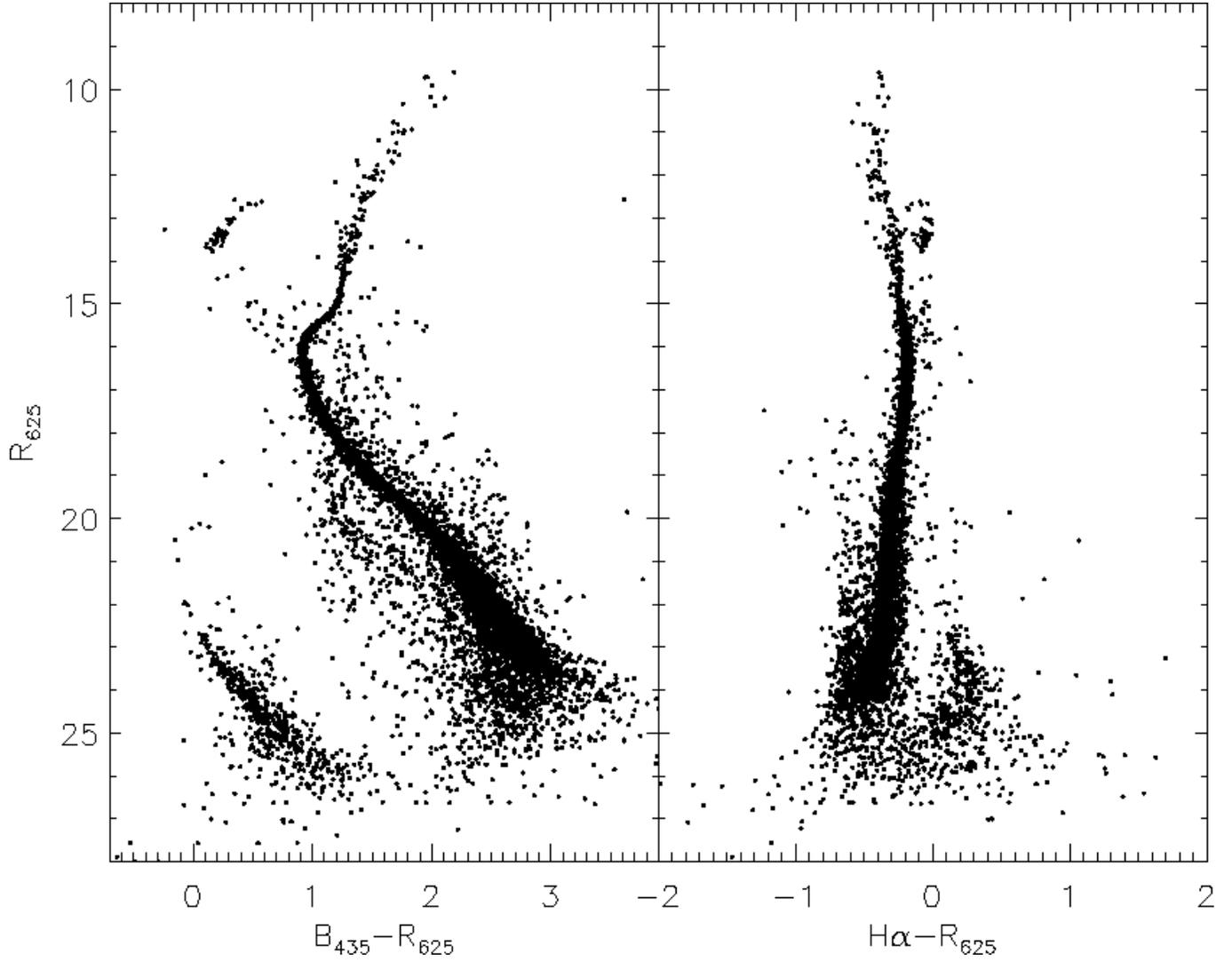}
\caption[] 
{Color-magnitude diagrams for 14,976 stars in the central regions of
NGC~6397.  The white dwarfs on the bottom left side of the \R625\ vs.\
\br\ diagram (left panel) are the focus of this paper.  In \hr\ vs.\
\R625\ these stars appear to the right of the main sequence,
indicating strong \ha\ absorption lines.  
}
\label{fig:CMD_full} 
\end{center} \end{figure}

\clearpage
\begin{figure}[h!] \begin{center} 
\includegraphics[angle=90, scale=0.85]{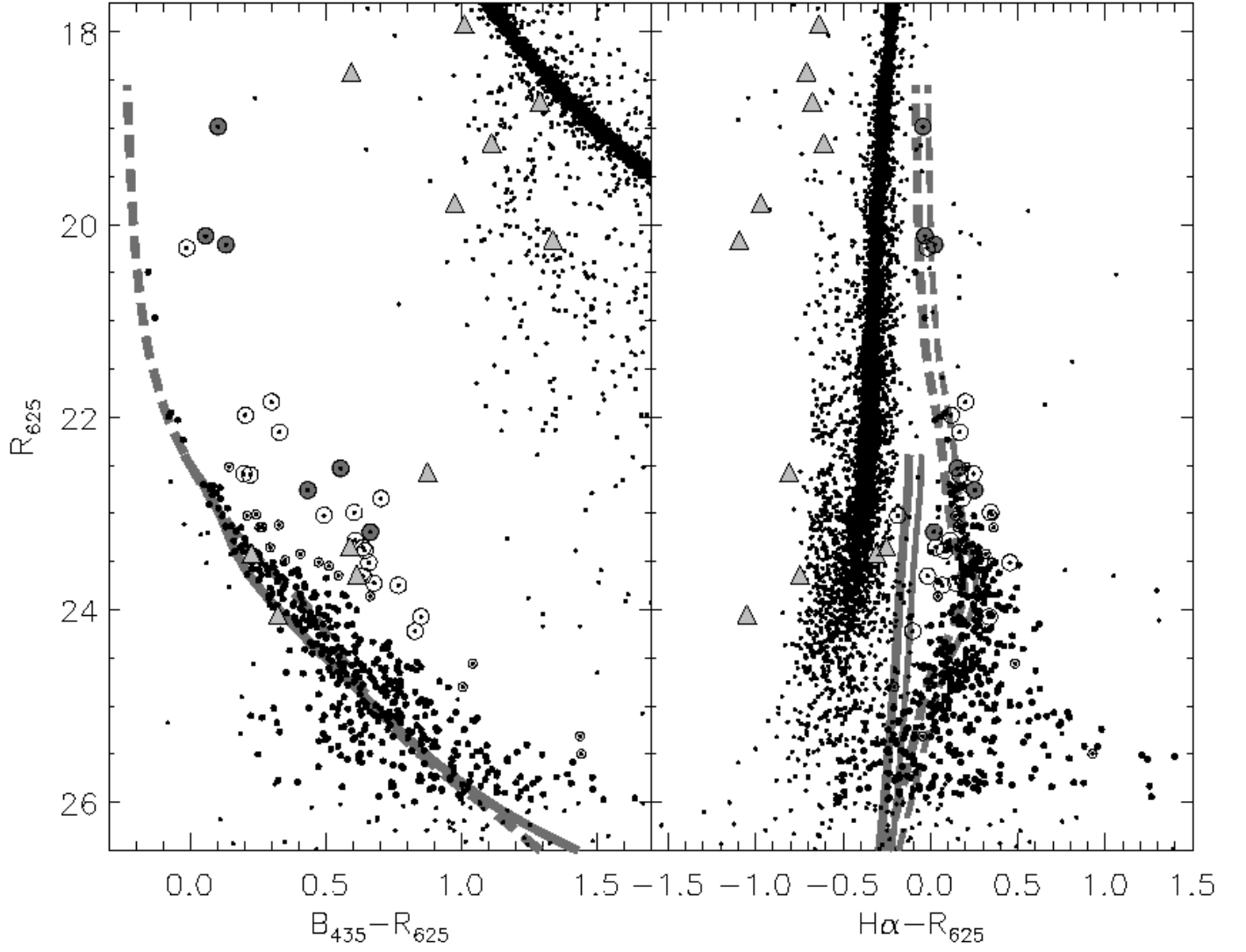}
\caption[] 
{White dwarfs in NGC~6397.  The 24 stars that we identify as the best
He~WD candidates are marked with large open circles.  These include 6
previously known candidates which are shaded grey.  Marginal He~WD
candidates (another 17 stars) are marked with small open circles.
Grey triangles mark the 11 known cataclysmic variables in the cluster.
Stars selected as probable CO~WDs are shown as large solid dots.  All
other stars are shown as small dots.  The thick dashed and solid lines
in both panels represent 0.53\msun\ DA and DB WD cooling tracks,
respectively, shown at the best-fit distance found in \S 5.  In the
\hr\ vs.\ \R625\ CMD, the same pair of tracks is also shown shifted
0.07 magnitudes to the right (thin lines) to better align the DA track
with the observed WDs (see \S 4).}
\label{fig:CMD_zoom_candidates}
\end{center} \end{figure}

\clearpage 
\begin{figure}[h!] \begin{center} 
\includegraphics[angle=90, scale=0.85]{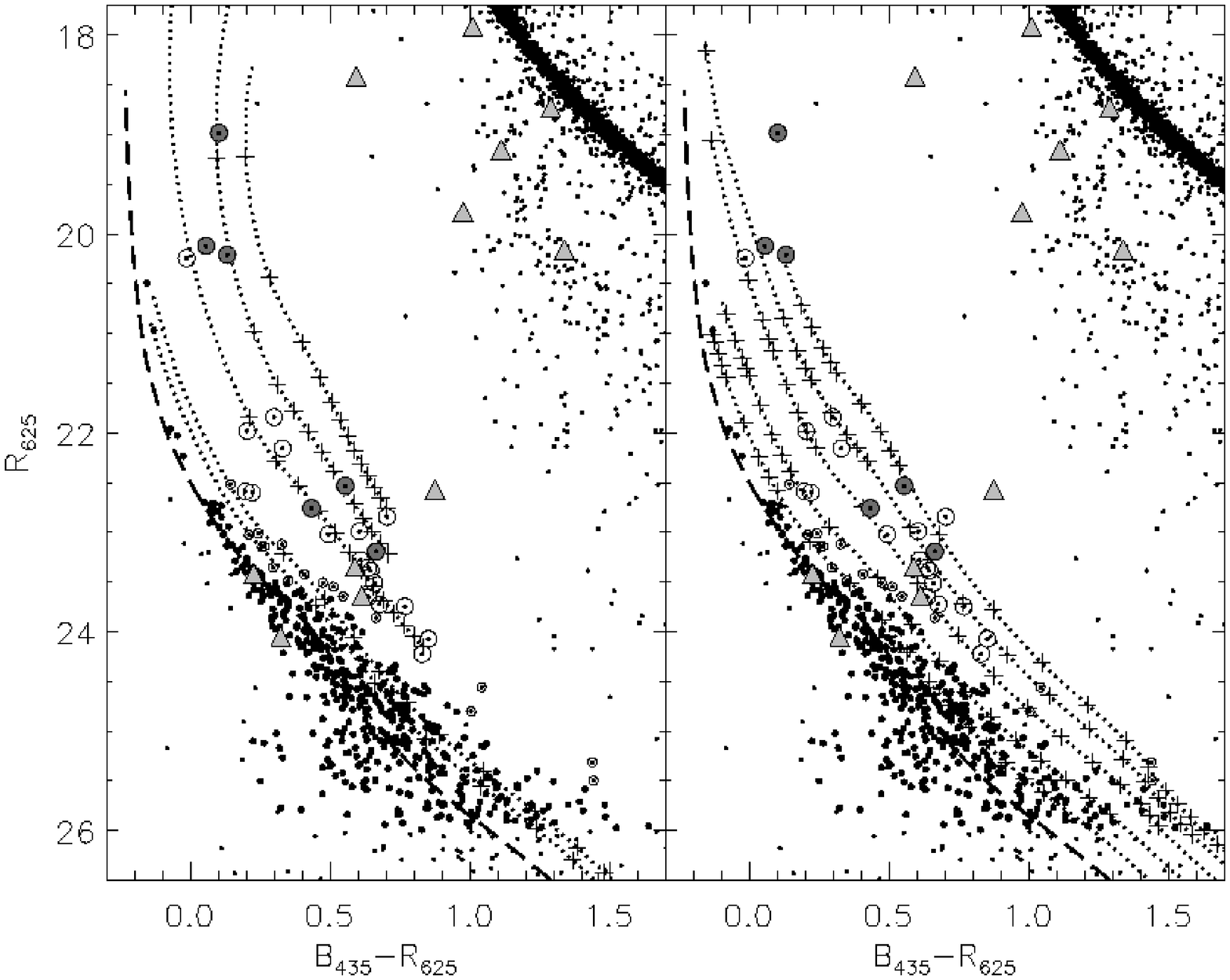}
\caption[] 
{Same as Fig.~3 but with He~WD cooling tracks from low-metallicity
progenitors (Z=0.0002) overlaid (dotted lines) \citep{sar02}.
The left and right panels show models with thick and thin hydrogen envelopes,
respectively (see \S 6.1).  In both panels, tracks are for He~WDs with
masses of 0.175\msun, 0.20\msun, 0.25\msun, 0.35\msun, and 0.45\msun\
(top to bottom).  In the left panel (thick envelopes), the ``+'' signs
indicate ages from $1-13$ Gyr in 1~Gyr intervals.  In the right panel
(thin envelopes), the ``+'' signs indicate ages of 0.001, 0.0025,
0.005, 0.0075, 0.01, 0.025, 0.05, 0.075, 0.1, 0.25, 0.5, 0.75, 1.0,
1.25, 1.5, 1.75, 2.0, 2.25, 2.5, and 2.75 Gyr.}
\label{fig:CMD_zoom_hewdtracks}
\end{center} \end{figure}

\clearpage
\begin{figure}[h!] \begin{center} 
\includegraphics[angle=90, scale=0.80]{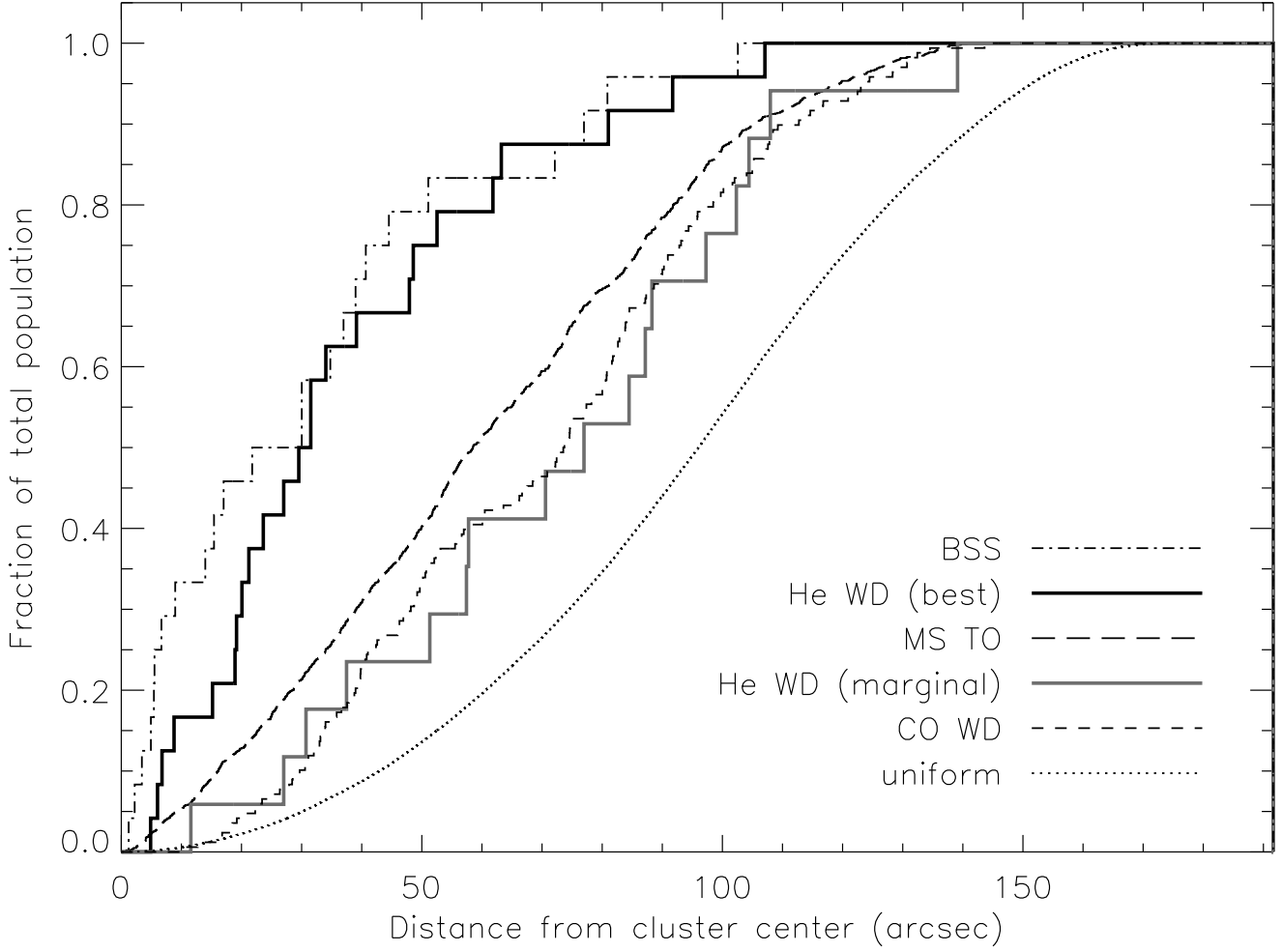}
\caption[] 
{Cumulative radial distributions of various populations of stars in
NGC~6397 and for a hypothetical uniformly distributed population.
These distributions are used for the Kolmogorov-Smirnov tests
described in the text (see \S 6.1).  The dark solid line shows the distribution of
the 24 best He~WD candidates; the solid grey line is for the 17
marginal candidates.  These can be compared with the distribution
expected if the stars were unrelated to the cluster (dotted line) and with the distributions of
168 CO~WDs (short-dash line), 1416 turnoff stars (long-dash line), and
24 blue stragglers (dash-dot line).}
\label{fig:KStest_all} 
\end{center} \end{figure}

\clearpage
\begin{figure}[h!] \begin{center} 
\includegraphics[angle=90, scale=0.80]{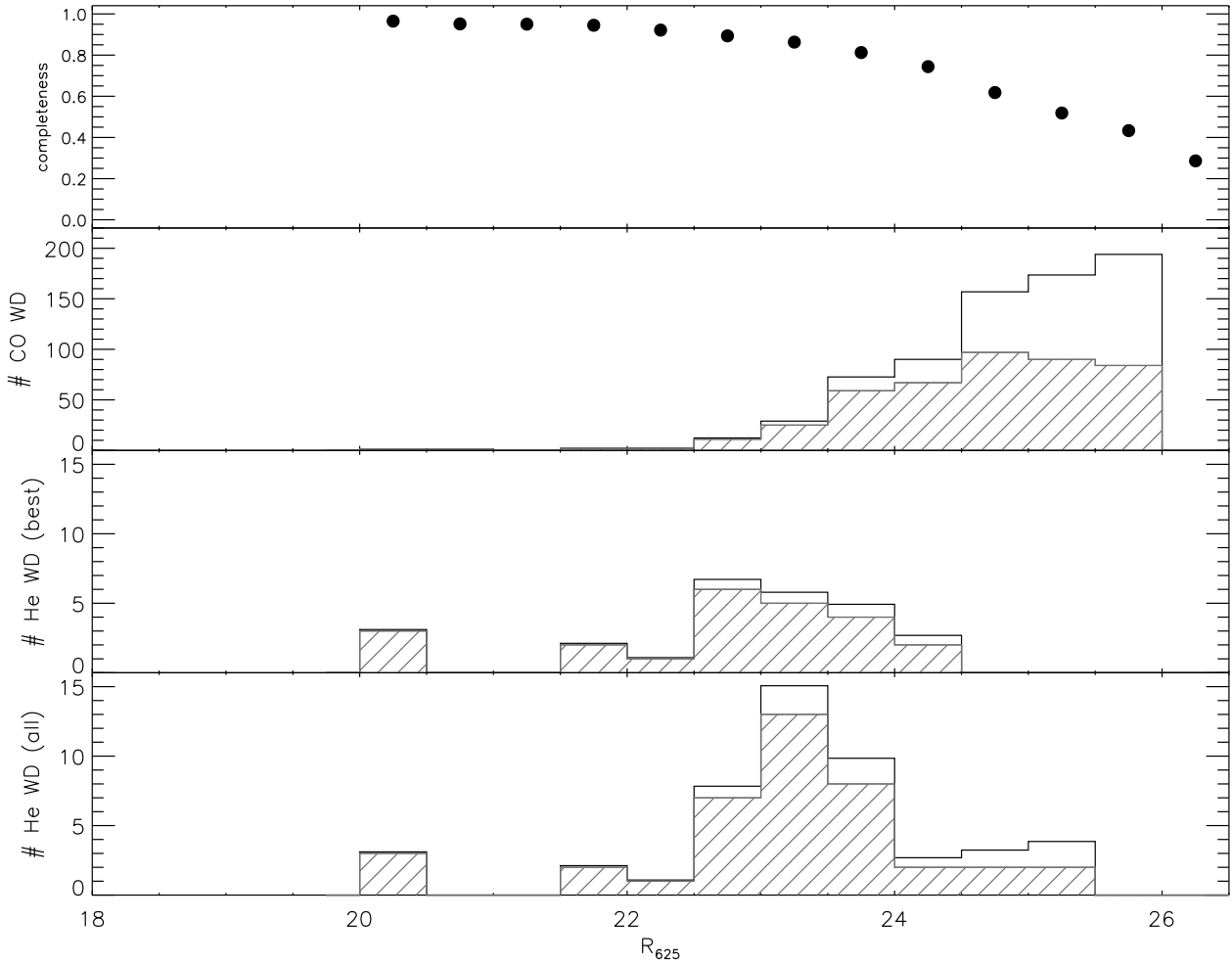}
\caption[] 
{Results of artificial star tests.
The top panel shows the fraction of stars recovered in artificial
star tests for stars on the WD sequence as a function of \R625\
magnitude.  We used these completeness statistics to infer the numbers
of CO~WDs (panel 2) and He~WDs (best 24 only: panel 3; all: panel 4)
in the cluster down to \R625 = 26.  Shaded histograms indicate
numbers observed; white histograms include the completeness
corrections.  From the two bottom panels we conclude that, in contrast
to the CO~WD sequence, the He~WD sequence ends before the magnitude
limit is reached.}
\label{fig:completeness} 
\end{center} \end{figure}

\clearpage
\begin{figure}[ht!] \begin{center} 
\includegraphics[angle=90, scale=0.85]{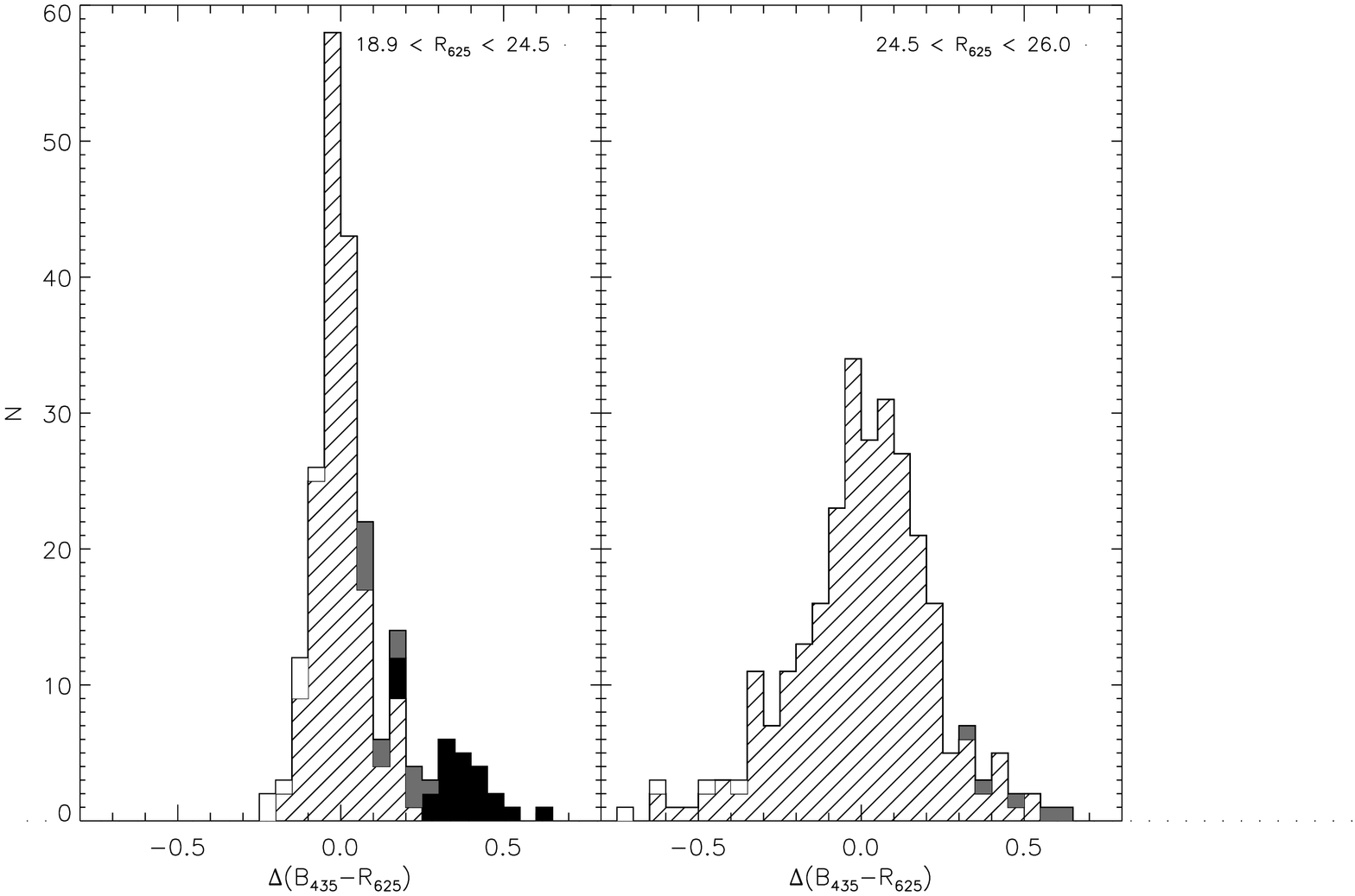}
\caption[] 
{Color distribution of CO~WDs and He~WD candidates, measured
relative to the color of the CO~WD cooling track shown in Fig.~3.  The
left panel shows stars with $18.9 \leq$ \R625\ $< 24.5$; the right
panel shows stars with $24.5 \leq$ \R625\ $< 26$.  Diagonal-line shading
denotes CO~WDs.  The best He~WD candidates are shown in black and the
marginal candidates in grey.  Stars not assigned to either class are
white.  Among the brighter stars (left panel) the distribution is
bimodal, with the He~WDs forming a distinct second peak.  At fainter
magnitudes (right panel), the broadening of the CO~WD sequence due to
increased measurement uncertainties makes it more difficult to discern
He~WDs.  However, there is no sign of any excess of stars on the right
side as compared to the left.}
\label{fig:color_distance}
\end{center} \end{figure}


\clearpage
\begin{deluxetable}{rcccrrrr}
\tabletypesize{\scriptsize} 
\tablewidth{0pt} 
\tablecaption{Helium-core White Dwarf Candidates. \label{table:hewdcandidates} } 
\tablehead{ 
ID & x, y & R.A. & Dec. & $\Delta$r\tablenotemark{1}  & $R_{625}$ & $B_{435}$$-$$R_{625}$ & $H\alpha$$-$$R_{625}$
}
\startdata
1\tablenotemark{2}  &  3147, 2683  &   17 40 40.659    &  $-$53 40 19.85  &  15.21  &  18.98  &  0.10  &  $-$0.04 \\  
2\tablenotemark{2}  &  3073, 2540  &   17 40 41.076    &  $-$53 40 26.98  &  8.82  &  20.12  &  0.06  &  $-$0.03 \\  
3\tablenotemark{2}  &  2780, 2529  &   17 40 42.722    &  $-$53 40 27.53  &  6.09  &  20.21  &  0.13  &  0.02 \\  
4  &  3208, 2146  &   17 40 40.314    &  $-$53 40 46.68  &  23.66  &  20.24  &  $-$0.01  &  $-$0.02 \\  
5  &  2511, 3382  &   17 40 44.237    &  $-$53 39 44.89  &  47.96  &  21.84  &  0.30  &  0.20 \\  
6  &  1650, 1160  &   17 40 49.083    &  $-$53 41 35.97  &  91.79  &  21.98  &  0.20  &  0.12 \\  
7  &  2349, 3401  &   17 40 45.145    &  $-$53 39 43.95  &  52.55  &  22.16  &  0.33  &  0.17 \\  
8\tablenotemark{3}  &  2872, 3046  &   17 40 42.206    &  $-$53 40 01.67  &  27.09  &  22.53  &  0.56  &  0.15 \\  
9  &  1646, 2328  &   17 40 49.103    &  $-$53 40 37.57  &  63.29  &  22.59  &  0.20  &  0.25 \\  
10\tablenotemark{3} &  2942, 2417  &   17 40 41.813    &  $-$53 40 33.16  &  4.91  &  22.76  &  0.43  &  0.25 \\  
11 &  2383, 2868  &   17 40 44.957    &  $-$53 40 10.57  &  31.58  &  22.60  &  0.22  &  0.17 \\  
12 &  2219, 2489  &   17 40 45.882    &  $-$53 40 29.56  &  34.07  &  22.84  &  0.70  &  0.18 \\  
13 &  3391, 2834  &   17 40 39.285    &  $-$53 40 12.30  &  29.55  &  22.99  &  0.61  &  0.34 \\  
14 &  4980, 1989  &   17 40 30.344    &  $-$53 40 54.51  &  107.14  &  23.02  &  0.49  &  $-$0.19 \\  
15\tablenotemark{3} &  2875, 2371  &   17 40 42.189    &  $-$53 40 35.45  &  6.84  &  23.19  &  0.67  &  0.02 \\  
16 &  2652, 2823  &   17 40 43.444    &  $-$53 40 12.84  &  20.14  &  23.28  &  0.61  &  0.11 \\  
17 &  2940, 2128  &   17 40 41.823    &  $-$53 40 47.57  &  18.96  &  23.36  &  0.64  &  0.03 \\  
18 &  1639, 1485  &   17 40 49.144    &  $-$53 41 19.72  &  81.08  &  23.39  &  0.65  &  0.08 \\  
19 &  2481, 2430  &   17 40 44.406    &  $-$53 40 32.52  &  21.28  &  23.51  &  0.66  &  0.45 \\  
20 &  2182, 3162  &   17 40 46.085    &  $-$53 39 55.91  &  48.61  &  23.65  &  0.64  &  $-$0.02 \\  
21 &  2449, 1352  &   17 40 44.585    &  $-$53 41 26.38  &  61.90  &  23.72  &  0.68  &  0.15 \\  
22 &  2912, 2122  &   17 40 41.982    &  $-$53 40 47.92  &  19.20  &  23.75  &  0.77  &  0.07 \\  
23 &  3529, 2567  &   17 40 38.511    &  $-$53 40 25.63  &  31.59  &  24.07  &  0.85  &  0.34 \\  
24 &  3668, 2656  &   17 40 37.725    &  $-$53 40 21.17  &  39.16  &  24.22  &  0.83  &  $-$0.10 \\  
\hline
25  &  1810, 2869  &   17 40 48.183    &  $-$53 40 10.54  &  57.46  &  22.52  &  0.14  &  0.20 \\  
26  &  2852, 2278  &   17 40 42.317    &  $-$53 40 40.10  &  11.62  &  23.01  &  0.24  &  0.35 \\  
27  &  1447, 3511  &   17 40 50.218    &  $-$53 39 38.42  &  88.33  &  23.03  &  0.21  &  0.14 \\  
28  &  1174, 4690  &   17 40 51.751    &  $-$53 38 39.49  &  139.17  &  23.12  &  0.33  &  0.17 \\  
29  &  3404, 2152  &   17 40 39.212    &  $-$53 40 46.39  &  30.78  &  23.14  &  0.27  &  0.16 \\  
30  &  1885, 2663  &   17 40 47.760    &  $-$53 40 20.84  &  51.35  &  23.15  &  0.26  &  0.36 \\  
31  &  4196, 4234  &   17 40 34.759    &  $-$53 39 02.31  &  108.01  &  23.36  &  0.30  &  0.16 \\  
32  &  2190, 2749  &   17 40 46.045    &  $-$53 40 16.54  &  37.54  &  23.42  &  0.41  &  0.32 \\  
33  &  2511, 1147  &   17 40 44.239    &  $-$53 41 36.63  &  70.64  &  23.49  &  0.35  &  0.28 \\  
34  &  3363, 2786  &   17 40 39.443    &  $-$53 40 14.70  &  27.07  &  23.51  &  0.47  &  0.35 \\  
35  &  4813, 3345  &   17 40 31.287    &  $-$53 39 46.71  &  104.46  &  23.55  &  0.51  &  0.23 \\  
36  &  2762, 4447  &   17 40 42.825    &  $-$53 38 51.65  &  97.32  &  23.64  &  0.55  &  0.30 \\  
37  &  1529, 4027  &   17 40 49.758    &  $-$53 39 12.65  &  102.38  &  23.86  &  0.66  &  0.04 \\  
38  &  4275, 3578  &   17 40 34.314    &  $-$53 39 35.07  &  87.20  &  24.56  &  1.04  &  0.48 \\  
39  &  1899, 1926  &   17 40 47.683    &  $-$53 40 57.70  &  57.83  &  24.80  &  1.01  &  $-$0.21 \\
40  &  2333, 3938  &   17 40 45.236    &  $-$53 39 17.07  &  77.05  &  25.31  &  1.44  &  $-$0.05 \\
41  &  3438, 903  &   17 40 39.021    &  $-$53 41 48.85  &  84.52  &  25.50  &  1.44  &  0.93 \\
\enddata
\tablenotetext{1}
{Offset in arcseconds from the cluster center.  For the center we
adopted 17$^\mathrm{h}$ 40$^\mathrm{m}$ 42\spt049, $-$53$^\circ$
40$\arcmin$ 28\secspt72 from \citet{sosin97}, which corresponds to (x,
y) = (2900, 2506) in the mosaic.}
\tablenotetext{2}{ID=1, 2, and 3 correspond to NF1, NF2, and NF3 identified by \citet{cgc98}, respectively. }
\tablenotetext{3}{ID=8, 10, and 15 correspond to PC4, PC5, and PC6 identified by \citet{tge01}, respectively. }
\end{deluxetable}

\end{document}